\documentclass[prl,twocolumn,superscriptaddress,citeautoscript,showpacs,amsart,longbibliography]{revtex4}

\usepackage{graphicx}
\usepackage{appendix}
\usepackage{multirow}
\usepackage{color}
\usepackage{bm}
\usepackage{times}
\usepackage{amsmath,bm,amsfonts}
\usepackage{dcolumn}
\usepackage{graphicx}
\usepackage{latexsym}
\usepackage{BOONDOX-cal}

\usepackage{ulem} 
\usepackage{braket}
\usepackage{mathtools}

\usepackage{cancel}

\begin{document}


\title{Thermal Hall effect, spin Nernst effect, and spin density induced by thermal gradient in \\ collinear ferrimagnets from magnon-phonon interaction}

\author{Sungjoon \surname{Park}}

 \affiliation{Department of Physics and Astronomy, Seoul National University, Seoul 08826, Korea}

\affiliation{Center for Correlated Electron Systems, Institute for Basic Science (IBS), Seoul 08826, Korea}

\affiliation{Center for Theoretical Physics (CTP), Seoul National University, Seoul 08826, Korea}

\author{Naoto \surname{Nagaosa}}

\affiliation{Department of Applied Physics, The University of Tokyo, Bunkyo, Tokyo 113-8656, Japan}

\affiliation{RIKEN Center for Emergent Matter Science (CEMS), Wako, Saitama 351-0198, Japan}

\author{Bohm-Jung \surname{Yang}}
\email{bjyang@snu.ac.kr}
 \affiliation{Department of Physics and Astronomy, Seoul National University, Seoul 08826, Korea}

\affiliation{Center for Correlated Electron Systems, Institute for Basic Science (IBS), Seoul 08826, Korea}

\affiliation{Center for Theoretical Physics (CTP), Seoul National University, Seoul 08826, Korea}

\date{\today}

\begin{abstract}
We theoretically study the intrinsic thermal Hall and spin Nernst effect in collinear ferrimagnets on a honeycomb lattice with broken inversion symmetry.
The broken inversion symmetry allows in-plane Dzyaloshinskii-Moriya interaction between the nearest neighbors, which does not affect the magnon spectrum in the linear spin wave theory.
However, the Dzyaloshinskii-Moriya interaction can induce large Berry curvature in the magnetoelastic excitation spectrum through the magnon-phonon interaction to produce thermal Hall current.
Furthermore, we find that the magnetoelastic excitations transport spin, which is inherited from the magnon bands.
Therefore, the thermal Hall current is accompanied by spin Nernst current.
Because the magnon-phonon interaction does not conserve the spin, we also study the spin density induced by thermal gradient in the presence of magnon-phonon interaction.
We find that the intrinsic part of the spin density shows no asymmetric spin accumulation near the boundary of the system having a stripe geometry.
However, because of the magnon-phonon interaction, we find nonzero total spin density in the system having armchair edges.
The extrinsic part of the spin density, on the other hand, shows asymmetric spin accumulation near the boundary for both armchair and zigzag edges because of the magnon-phonon interaction. In addition, we find nonzero total spin density in the system having zigzag edge.
\end{abstract}

\maketitle

Recently, there is a growing interest in the theoretical and experimental studies of the thermal Hall conductivity of insulators arising from charge neutral quasiparticles such as magnons \cite{katsura2010theory,onose2010observation,matsumoto2011theoretical,matsumoto2011rotational,matsumoto2014thermal} and phonons \cite{strohm2005phenomenological,sheng2006theory,inyushkin2007phonon,kagan2008anomalous,zhang2010topological,qin2012berry,mori2014origin,saito2019berry}.
It was found that the intrinsic thermal Hall conductivity of bosonic excitations is a manifestation of the Berry curvature of their wave function~\cite{matsumoto2011theoretical,zhang2010topological,qin2012berry}. 
This led to the proposal of topological bosonic band structure with nonzero Chern number that shows thermal Hall effect \cite{shindou2013topological,zhang2013topological,zhang2010topological}.
Moreover, when the bosonic excitation carries spin quantum number, analogues of quantum spin Hall insulators were proposed \cite{nakata2017magnonic,lee2018magnonic,kondo2019z} that show spin Nernst effect \cite{cheng2016spin,zyuzin2016magnon,kovalev2016spin,zhang2018spin,mook2019spin}.

In these theoretical studies, it was assumed that the low energy excitations are either magnons or phonons. 
However, it was realized that large Berry curvature can be induced in the anticrossing regions between the magnon and phonon bands through the magnon-phonon interaction (MPI)  \cite{takahashi2016berry}.
Using this idea, it was shown that in a non-collinear antiferromagnet, magnon and phonon bands with zero Chern numbers can hybridize through the MPI originating from magnetostriction to form magnetoelastic bands with nonzero Chern numbers, thus enhancing the thermal Hall conductivity~\cite{park2019topological}.
Similar ideas were also put forward in Refs.~\cite{go2019topological,zhang2019thermal} in the case of ferromagnetic insulators, where the authors proposed that the thermal Hall conductivity can be produced solely by the MPI originating from magnetic anisotropy \cite{go2019topological} or Dzyaloshinskii-Moriya (DM) interaction ~\cite{zhang2019thermal}. However, the study of spin dependent response from magnetoelastic excitations is still missing.

In this work, we propose that not only the intrinsic thermal Hall current, but also the intrinsic spin Nernst current can be produced entirely by the MPI in a collinear ferrimagnet with honeycomb lattice, in which the inversion symmetry between the nearest neighbors is broken. 
As is well known \cite{keffer1953spin,cheng2016spin,zyuzin2016magnon}, collinear antiferromagnet on a honeycomb lattice with easy axis anisotropy supports two magnon bands carrying opposite spin quantum numbers. 
The same is also true for ferrimagnets, so that
we can split the energies of two magnon bands by applying an external magnetic field, which  also allows us to control the band crossing regions between the acoustic phonon bands and the spin up or down magnon band.
By considering the MPI arising from in-plane DM interaction between the nearest neighbors \cite{zhang2019thermal}, we find that the gaps between the magnon and phonon bands open to produce anticrossing regions with large Berry curvature.
Because the magnon and phonon bands show no thermal Hall response without MPI, the thermal Hall conductivity in our model arises solely from the MPI.
Furthermore, because the magnon bands carry spin, the thermal Hall current carried by magnetoelastic excitations is spin polarized, and therefore, it is accompanied by spin Nernst current. Because the magnon bands show no spin Nernst response without MPI, the spin Nernst current also arises purely from MPI.

Because the MPI violates the spin conservation, we reexamine the derivation of the spin Nernst coefficient using Luttinger's method of gravitational potential \cite{luttinger1964theory,zyuzin2016magnon,matsumoto2011rotational,matsumoto2014thermal} and compare the result with spin density induced by thermal gradient. 
We find that the intrinsic contribution to the spin density shows no asymmetric boundary spin accumulation for the system having stripe geometry.
However, the extrinsic contribution to the spin density, which relies on finite quasiparticle lifetime, reflects the spin Nernst current through asymmetric  spin accumulation \cite{murakami2005intrinsic}. 
Moreover, nonzero total spin expectation value is induced by the intrinsic (extrinsic) mechanism in the system having armchair (zigzag) edges, but not for zigzag (armchair) edges.
Finally, we discuss the relevance of our model to the thermal Hall conductivity measured in $\textrm{Fe}_2\textrm{Mo}_3\textrm{O}_8$ \cite{ideue2017giant}.

\begin{figure}[t]
\centering
\includegraphics[width=8.5cm]{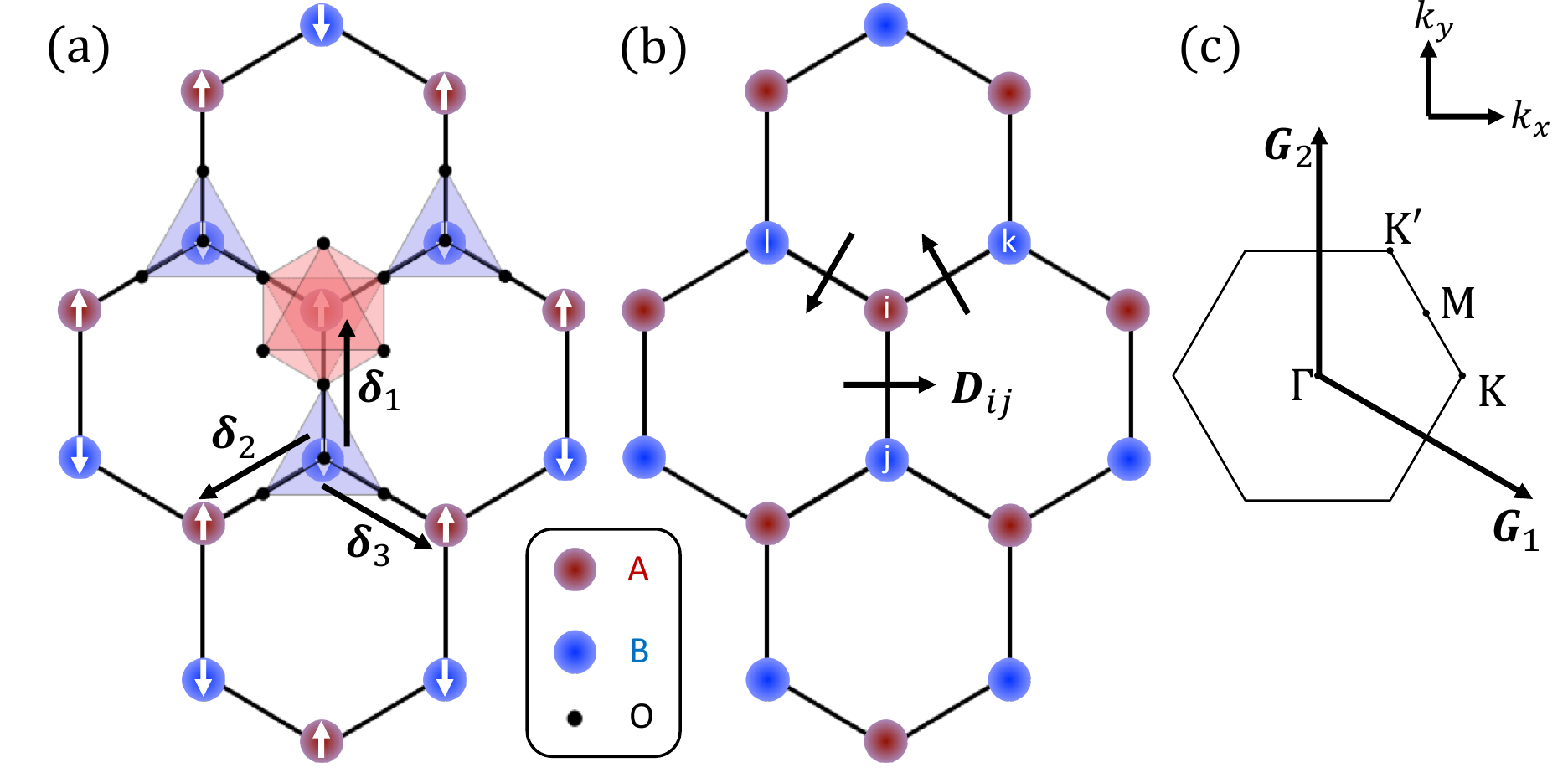}
\caption{(a)  Magnetic order on a honeycomb lattice with broken inversion symmetry between $A$ and $B$ sites, which are surrounded respectively by octahedral and tetrahedral cages of oxygen atoms. The up spins point out of the plane, and the down spins point into the plane.
(b) Simplified lattice model where we indicate the directions of the in-plane DM vectors $\bm{D}_{ij}$, where $i$ is fixed to be one of the $A$ sites and $j$ is one of the three nearest $B$ sites.
(c) The Brillouin zone and the high symmetry momenta. }
\label{fig.lattice}
\end{figure}

\textit{Model.|} 
Our model is motivated by $\textrm{Fe}_2\textrm{Mo}_3\textrm{O}_8$, which consists of Fe-O layers separated by sheets of $\textrm{Mo}^{4+}$ \cite{le1982structure,wang2015unveiling}.
The $\textrm{Fe}^{2+}$ ions in the Fe-O layer form a honeycomb-like lattice as in Fig.~\ref{fig.lattice} (a), where the $\textrm{Fe}^{2+}$ ions at $A$ and $B$ sites are displaced vertically with respect to each other.
Because the $A$ ($B$) sites are surrounded by octahedral (tetrahedral) cage of oxygen atoms, the inversion symmetry between the nearest neighbors is broken. 
However, the mirror symmetry $\mathcal{M}_x$: $(x,y)$ $\rightarrow$ $(-x,y)$ about the center of a hexagon is retained.

For our model, we simplify the lattice as shown in Fig.~\ref{fig.lattice} (b), where we keep only the magnetic ions forming a two-dimensional honeycomb lattice.
We consider the spin Hamiltonian given by
\begin{equation}
\mathcal{H}_m=\mathcal{H}_{J_1} +\mathcal{H}_{J_2}+\mathcal{H}_D+\mathcal{H}_{\alpha}+\mathcal{H}_{H}. \label{eq.magnon_hamiltonian}
\end{equation}
Here, $\mathcal{H}_{J_1}=J_1\sum_{\langle ij \rangle }  \bm{S}_i\cdot \bm{S}_j$ with $J_1>0$ is the antiferromagnetic nearest-neighbor Heisenberg interaction, and
$\mathcal{H}_{J_2}=J_2^A \sum_{\langle \langle ij \rangle \rangle _A}\bm{S}_i\cdot \bm{S}_j+J_2^B \sum_{\langle \langle ij \rangle \rangle _B}\bm{S}_i\cdot \bm{S}_j$ is the ferromagnetic next-nearest-neighbor Heisenberg interaction between the $A$ sites ($J_2^A<0$) and $B$ sites ($J_2^B<0$). 
 To reflect the broken inversion symmetry that relates $A$ and $B$ sites, we put $J_2^A\neq J_2^B$ and include the nearest-neighbor DM interaction
 $\mathcal{H}_D=\sum_{\langle ij \rangle} \bm{D}_{ij} \cdot [\bm{S}_i \times \bm{S}_j]$, where the direction of $\bm{D}_{ij}$ is indicated in Fig.~\ref{fig.lattice}.
We note, however, that the DM interaction does not contribute to the magnon spectrum because the magnetic ordering direction, normal to the honeycomb plane, is perpendicular to the DM vector.
Finally, $\mathcal{H}_{\alpha}=\sum_{i} \alpha_i(S_i^z)^2$ with $\alpha_i<0$ is the easy-axis anisotropy ($\alpha_A\neq \alpha_B$) and $\mathcal{H}_{H}=\sum_{i}\mu_i \bm{S}_i\cdot \bm{H}$ is the Zeeman coupling to the external magnetic field applied parallel to the magnetic ordering direction ($\mu_A\neq \mu_B$).

\begin{figure}[t]
\centering
\includegraphics[width=8.5cm]{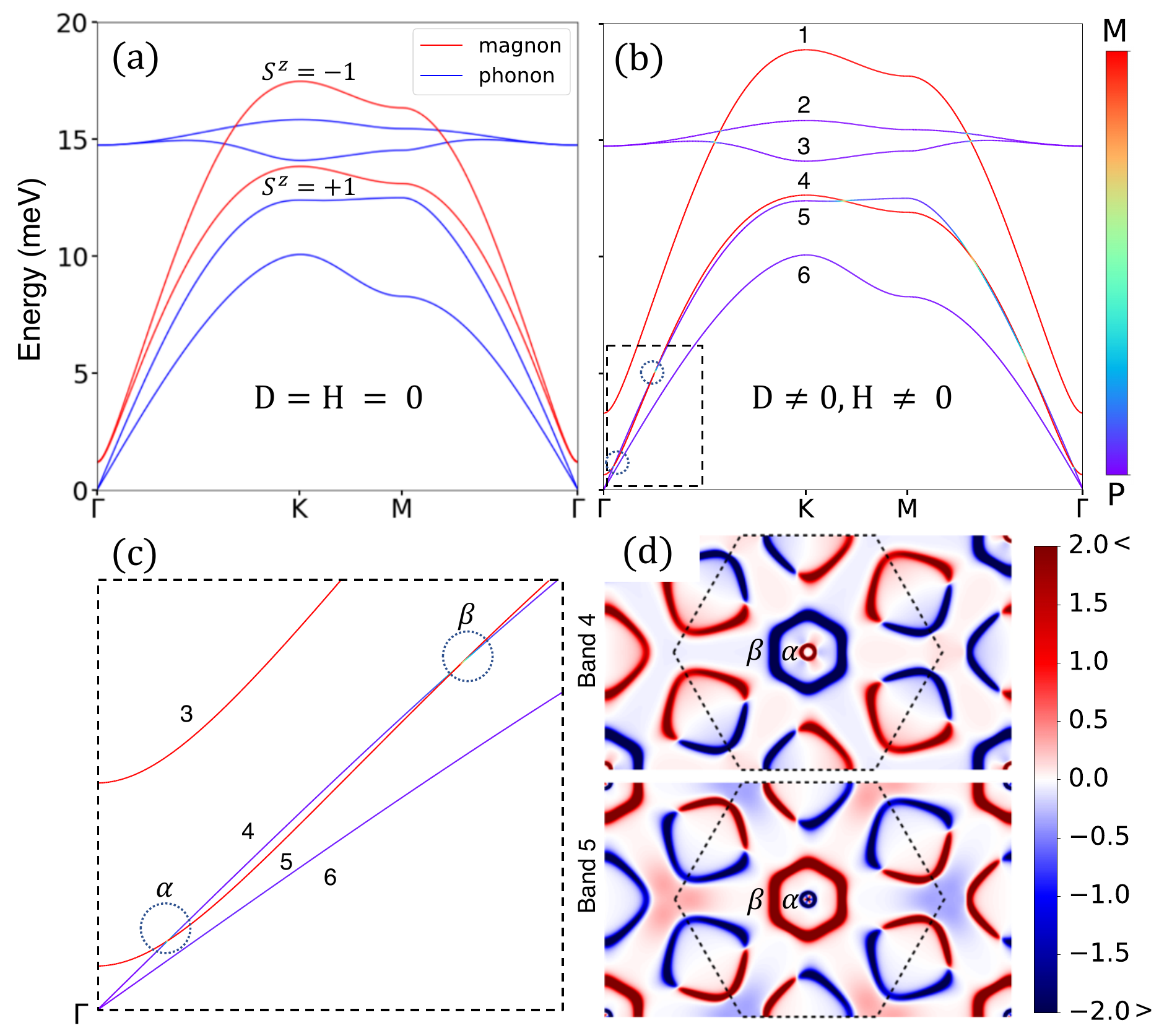}
\caption{ (a) Magnon and phonon spectra along the high symmetry lines with no MPI ($D=0$) and no magnetic field ($H=0$). The upper (lower) magnon band carries $-1$ ($+1$) spin. The parameters for the magnon Hamiltonian are $S=2$, $J_1=2$ meV, $J_2^A=-0.3$ meV, $J_2^B=-0.1$ meV, $\alpha_A=-0.02$ meV, and $\alpha_B=-0.01$ meV, and the parameters for the phonon Hamiltonian are $K_L=120$ $(\textrm{meV})^2$, $K_T=25$ $(\textrm{meV})^2$, $k_L^A=30$ $(\textrm{meV})^2$, $k_T^A=10$ $(\textrm{meV})^2$, $k_L^B=15$ $(\textrm{meV})^2$, $k_T^B=5$ $(\textrm{meV})^2$.
(b) Magnetoelastic excitation spectrum with MPI and magnetic field. The parameters relevant for the  MPI are $|\bm{D}_{ij}|=0.94$ meV, $l=5$ \AA ,  and $M_A=M_B=60$ amu, and the Zeeman interactions are $\mu_A H=-1.4$ meV and $\mu_B H=-1.2$ meV. The highest to lowest energy bands are labeled from 1 to 6.
The line color indicates the magnon and phonon content of the magnetoelastic modes. 
(c) Close-up view of the two anticrossing regions ($\alpha, \beta$) near the $\Gamma$ point, with a tiny gap at each crossing point. 
(d) Berry curvature density for energy bands 4 and 5. The boundary of the first Brillouin zone is indicated with dotted line.
}
\label{fig.hsl}
\end{figure}

The magnon Hamiltonian is obtained by writing $\bm{S}_i=\hat{\bm{x}}_iS_i^x+\hat{\bm{y}}_iS_i^y+\hat{\bm{z}}_iS_i^z$ where $\hat{\bm{x}}_i$, $\hat{\bm{y}}_i$, and $\hat{\bm{z}}_i$ are local orthogonal coordinates,  and introducing the Holstein-Primakoff operators $a_i$ and $a_i^\dagger$ as detailed in the Supplemental Material (SM) \cite{supplement}.
The magnon spectrum with $\bm{H}=0$ is shown in Fig.~\ref{fig.hsl} (a).
We find that the upper (lower) magnon band carries spin $-1(+1)$ by using the definition of the magnon spin operator $\mathcal{S}^z=\sum_{i} \textrm{sgn}(i) a^\dagger_i a_i $, where $\textrm{sgn}(i)=-1(+1)$ when $i$ is one of the A (B) sites. 
Note that although the DM interaction breaks the $\textrm{SO}(2)$ symmetry about the $z$ axis, it does not appear in the linear spin wave theory, and the magnon spin is conserved in this limit.

For the phonon Hamiltonian,  we consider a simple harmonic oscillator model of the form
\begin{equation}
\mathcal{H}_{p}=\frac{1}{2}\sum_{ij} \left[ \tilde{\bm{p}}^2_{i} \delta_{ij}+ \tilde{\bm{u}}_i K_{ij}(\bm{R}_i-\bm{R}_j)\tilde{\bm{u}}_{j}\right],\label{eq.phonon_H}
\end{equation}
where $\bm{R}_i$ is the position of the atom at site $i$ and $K_{ij}(\bm{R}_i-\bm{R}_j)$ are the spring constant matrices between the atoms at sites $i$ and $j$.
Note that we have absorbed the atomic mass $M_i$ by defining the rescaled displacement and momentum operators $\tilde{\bm{u}}_{i}=\sqrt{M_i}\bm{u}_{i}$ and $\tilde{\bm{p}}_i=\frac{\bm{p}_i}{\sqrt{M}_i}$, where $\bm{u}_i$ and $\bm{p}_i$ are the  displacement and momentum operators.
Let $K_L$ and $K_T$ be the nearest-neighbor longitudinal and transverse spring constants, respectively, and let $k_L^A$ and $k_T^A$ ($k_L^B$ and $k_T^B$) be the next-nearest-neighbor longitudinal and transverse spring constants between AA  (BB) sites.
These can be organized into spring constant matrices as discussed in the SM \cite{supplement}.
The resulting phonon spectrum is shown in Fig.~\ref{fig.hsl} (a).

Finally, we consider the MPI arising from the fluctuation of $\bm{D}_{ij} $when  the atoms deviate from their equilibrium positions~\cite{nomura2019phonon,zhang2019thermal,keffer1962moriya} by Taylor expanding $\bm{D}_{ij}$ in terms of $\frac{\bm{u}_i-\bm{u}_j}{l}$, where $l$ is the distance between nearest neighbors (the full expression of MPI is given in the SM). 
Note that because MPI mixes magnon and phonon, $\mathcal{S}^z$ is not conserved.
We show the magnetoelastic excitation spectrum with MPI in Fig.~\ref{fig.hsl} (b), where we also turn on the out-of-plane  magnetic field.
The energy bands with up and down spins respond oppositely to the magnetic field, so that with our choice of parameters (see Fig.~\ref{fig.hsl}), the energy of the spin up and down bands are lowered and raised, respectively.
This produces two overlapping regions near the $\Gamma$ point between the magnon and phonon bands, which hybridize because of the DM interaction. 
This is clarified in Fig.~\ref{fig.hsl} (c), where the two anticrossing regions between energy bands 4 and 5  near the $\Gamma$ point are indicated with dotted circles.
It is important to note that these two anticrossing regions correspond to the Berry curvature hotspots encircling the $\Gamma$ point in Fig.~\ref{fig.hsl} (d), which are crucial for the thermal Hall and spin Nernst effect.

\textit{Thermal Hall conductivity.|}
The thermal Hall conductivity $\kappa_{xy}$ is defined by the expression  $j^Q_{x}=-\kappa_{xy}\nabla_{y}T$, where $\bm{j}^Q$ is the heat current and $T$ is the temperature.
The semiclassical and linear response theories both yield~\cite{matsumoto2014thermal,zhang2016berry,park2019topological}  $\kappa_{xy}=-\frac{k_B^2 T}{V\hbar }\sum_{\bm{k}}\sum_{n=1}^{N} \left[c_2(g(E_{n,\bm{k}}))-\frac{\pi^2}{3}\right] \Omega_{n,\bm{k}}$, where the summation is over only the particle bands, $\Omega_{n,\bm{k}}$ is the Berry curvature, $c_2(x)=(1+x) (\ln\frac{1+x}{x})^2-(\ln x)^2-2\textrm{Li}_2(-x)$, and $\textrm{Li}_2$ is the polylogarithm function $\textrm{Li}_n$ for $n=2$.

Without MPI, the individual magnon and phonon bands satisfy the $\mathcal{M}_x \mathcal{C}_{2x}^S$ symmetry, which forces $\kappa_{xy}=0$ \cite{supplement}. Here, $\mathcal{M}_x$ is the mirror symmetry about the plane normal to the $x$ axis, which acts on both spin and position degrees of freedom. $\mathcal{C}_{2x}^S$ acts only on the spin degrees of freedom, and it rotates all of the spin about the $x$ axis by $\pi$ without changing their position.

\begin{figure}[t]
\centering
\includegraphics[width=8.5cm]{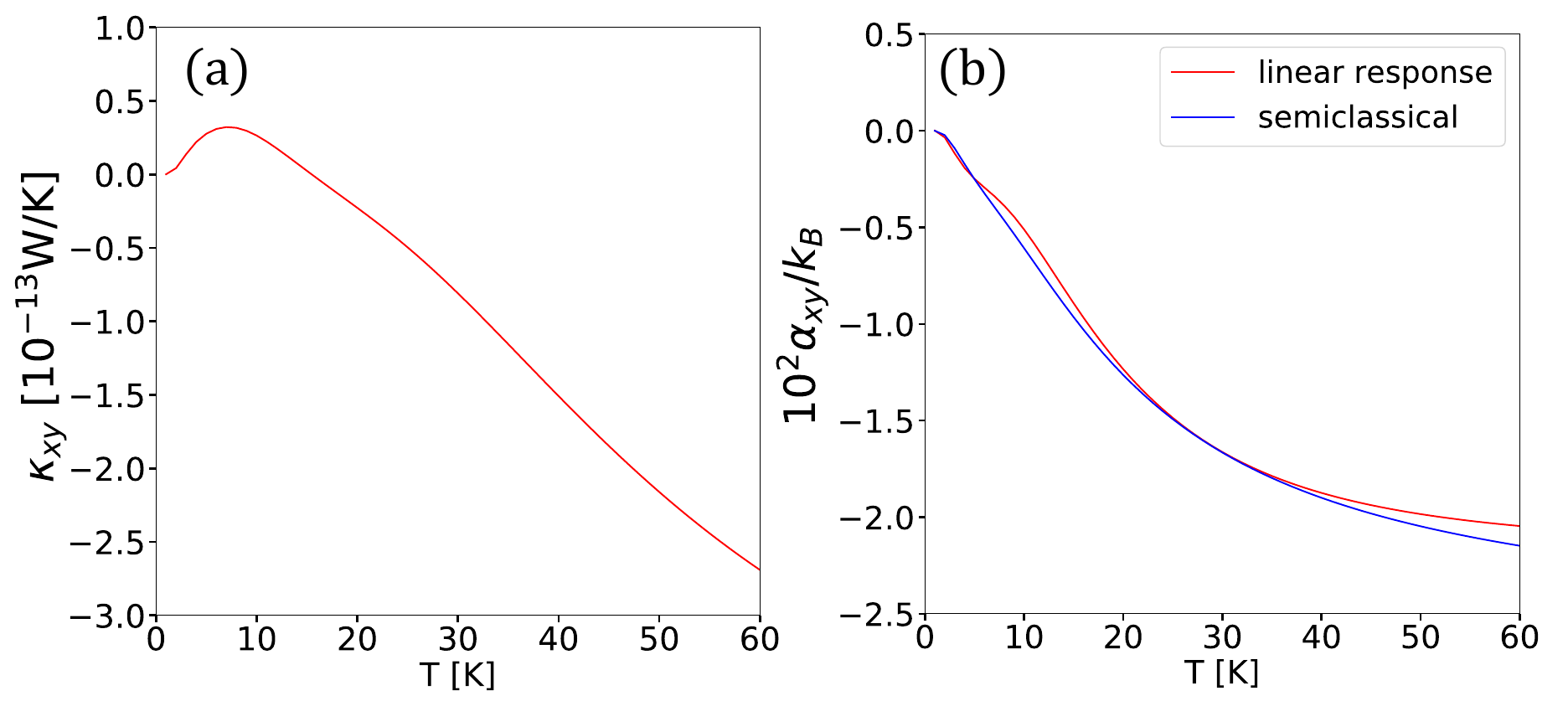}
\caption{ (a) Thermal Hall conductivity and (b) spin Nernst coefficient arising from MPI computed with the parameters used in Fig.~\ref{fig.hsl} (b). }
\label{fig.conductivity}
\end{figure}

Because DM interaction requires spin-orbit coupling, MPI breaks the $\mathcal{M}_x \mathcal{C}_{2x}^S$ symmetry, which decouples spin and orbital degrees of freedom. We therefore obtain nonzero $\kappa_{xy}$ as shown in Fig.~\ref{fig.conductivity} (a). 
Because the lowest three magnetoelastic bands in our model do not carry Chern numbers \cite{supplement}, the sign change in $\kappa_{xy}$ around $15$K cannot be explained by sign alternation of Chern numbers~\cite{hirschberger2015thermal,chisnell2015topological} between the magnetoelastic bands.
Instead, we notice that the two Berry curvature hotspots $\alpha, \beta$ near the $\Gamma$ point have opposite signs. 
Therefore, at low temperature, the smaller region ($\alpha$) with energy approximately $1$ meV and negative Berry curvature is the main contributor, so that $\kappa_{xy}<0$  (note that $k_B\times 10K\approx 0.86$ meV).
On the other hand, the larger region ($\beta$) with positive Berry curvature is located around $5$ meV, and therefore starts to contribute significantly at higher temperature to flips the sign of $\kappa_{xy}$.

\textit{Spin Nernst effect.|} 
The spin Nernst coefficient $\alpha^S_{xy}$ is defined from the expression for the spin current density $j^S_{x}=-\alpha^S_{xy}\nabla_y T$. 
When the spin is conserved, the semiclassical  \cite{cheng2016spin} and the linear response theory both yield $\alpha_{xy}^S=-\frac{k_B}{\hbar V}\sum_{\bm{k}}\sum_{n=1}^{N} \langle \mathcal{S}^z \rangle_{n,\bm{k}} \Omega_{n,\bm{k}} c_1(E_{n,\bm{k}}/k_BT)$, where $\langle \mathcal{S}^z \rangle_{n,\bm{k}}$ is the expectation value of $\mathcal{S}^z$, 
$c_1(x)=\int_{x}^\infty dx' x' \left(-\frac{d\rho(x')}{dx'}\right)=(1+\rho(x))\log(1+\rho(x))-\rho(x) \log \rho(x)$, and $\rho(x)=\frac{1}{e^{x}-1}$.

Similar to the case of $\kappa_{xy}$,  nonzero $\alpha_{xy}^S$ requires MPI to break the $\mathcal{M}_x  \mathcal{C}_{2x}^S$ symmetry \cite{supplement}.
We plot $\alpha_{xy}^S$ calculated using the semiclassical theory in Fig.~\ref{fig.conductivity} (b) (blue curve).
As can be seen, the behavior of the $\alpha_{xy}^S$ closely follows $\kappa_{xy}$.
This is because the low-energy magnetoelastic excitations  with large Berry curvature are mixtures of phonon and magnon with spin $+ 1$, so that thermal Hall current is accompanied by spin Nernst current.
However, in contrast to $\kappa_{xy}$, $\alpha_{xy}^S$ does not change sign near $T=15K$ because of the subtle distribution of $\langle \mathcal{S}^z \rangle_{n,\bm{k}}\Omega_{n,\bm{k}}$, which is analyzed in the SM \cite{supplement}.

Because it is not clear what approximations are made in the semiclassical approach, we also derive $\alpha_{xy}^S$ using the linear response theory
\cite{luttinger1964theory,smrcka1977transport,zyuzin2016magnon,matsumoto2011rotational,matsumoto2014thermal}. 
Writing the magnetoelastic Hamiltonian in bosonic BdG form \cite{supplement,park2019topological}, $\mathcal{H}_0=\frac{1}{2} \sum_{\bm{k}}\Psi^\dagger_{\bm{k}}H_{\bm{k}}\Psi_{\bm{k}}$, the energy eigenstates of $H_{\bm{k}}$ satisfy $\tau_z H_{\bm{k}}|n,\bm{k}\rangle=E_{\bm{k},n}|n,\bm{k}\rangle$ and $\langle m,\bm{k}| \tau_z |n,\bm{k} \rangle=(\tau_z)_{mn}$.
Here, $\Psi_{\bm{k}}$ contains the magnon and phonon fields and $\tau_z$ is the diagonal matrix with $(\tau_z)_{nn}=1$ ($-1$) in the particle (hole) space of the BdG Hamiltonian. 
When the MPI is small,  we find \cite{supplement} (omitting the $\bm{k}$ dependence for notational simplicity)
\begin{align}
\alpha^S_{\mu \nu}\approx \frac{k_B \hbar}{2V} \sum_{\bm{k},m\neq n} &\frac{i\langle n | w_\mu + u_\mu |m\rangle \langle m |v_\nu | n\rangle(\tau_z)_{mm} (\tau_z)_{nn}}{((\tau_z E)_{mm}-(\tau_z E)_{nn})^2}  \nonumber \\
& \times\left[c_1 \left(\frac{E_{mm}}{k_BT}\right)-c_1\left(\frac{E_{nn}}{k_BT}\right)\right],
\label{eq.spin_nernst}
\end{align}
where $\bm{v}=\frac{1}{\hbar}\frac{\partial H}{\partial \bm{k}}$, $\bm{w}=S^z \tau_z \bm{v}$, and $\bm{u}=\bm{v} \tau_z S^z$.

We show $\alpha_{xy}^S$ calculated using this approximation in Fig.~\ref{fig.conductivity} (b) (red curve).
We find that the behavior does not differ significantly from spin Nernst coefficient calculated using the semiclassical theory (blue curve).

\textit{Spin density.|} 
Because spin is not conserved by MPI  and edge spin accumulation is the experimentally measurable consequence of the spin current~\cite{sinova2006spin}, we study the spatial distribution of spin density as in electronic systems \cite{nomura2005edge,tse2005spin,sinova2006spin,go2018intrinsic,supplement}.
The spin density \footnote{We note that if we were interested in the magnetization, there should be a correction to the definition of magnetization resulting from the gravitational potential as noted in Refs.~\cite{shitade2019theory,shitade2019gravitomagnetoelectric} } induced by the thermal gradient at position $r$ is given by $\langle \delta\mathcal{S}^z(r) \rangle=\langle S^z(r)\rangle_{\textrm{neq}}-\langle S^z(r) \rangle_{\textrm{eq}}=-\zeta_\nu(r) \nabla_\nu T$.
Let us divide  $\zeta_\nu(r)=\zeta_\nu^{\textrm{in}}(r)+\zeta_\nu^{\textrm{ext}}(r)$ and study the two parts separately.
Here, the intrinsic part $\zeta_\nu^{\textrm{in}}(r)$ is independent of quasiparticle lifetime, while the extrinsic part $\zeta_\nu^{\textrm{ext}}(r)$ is approximately proportional to the quasiparticle lifetime.

\begin{figure}[t]
\centering
\includegraphics[width=8.5cm]{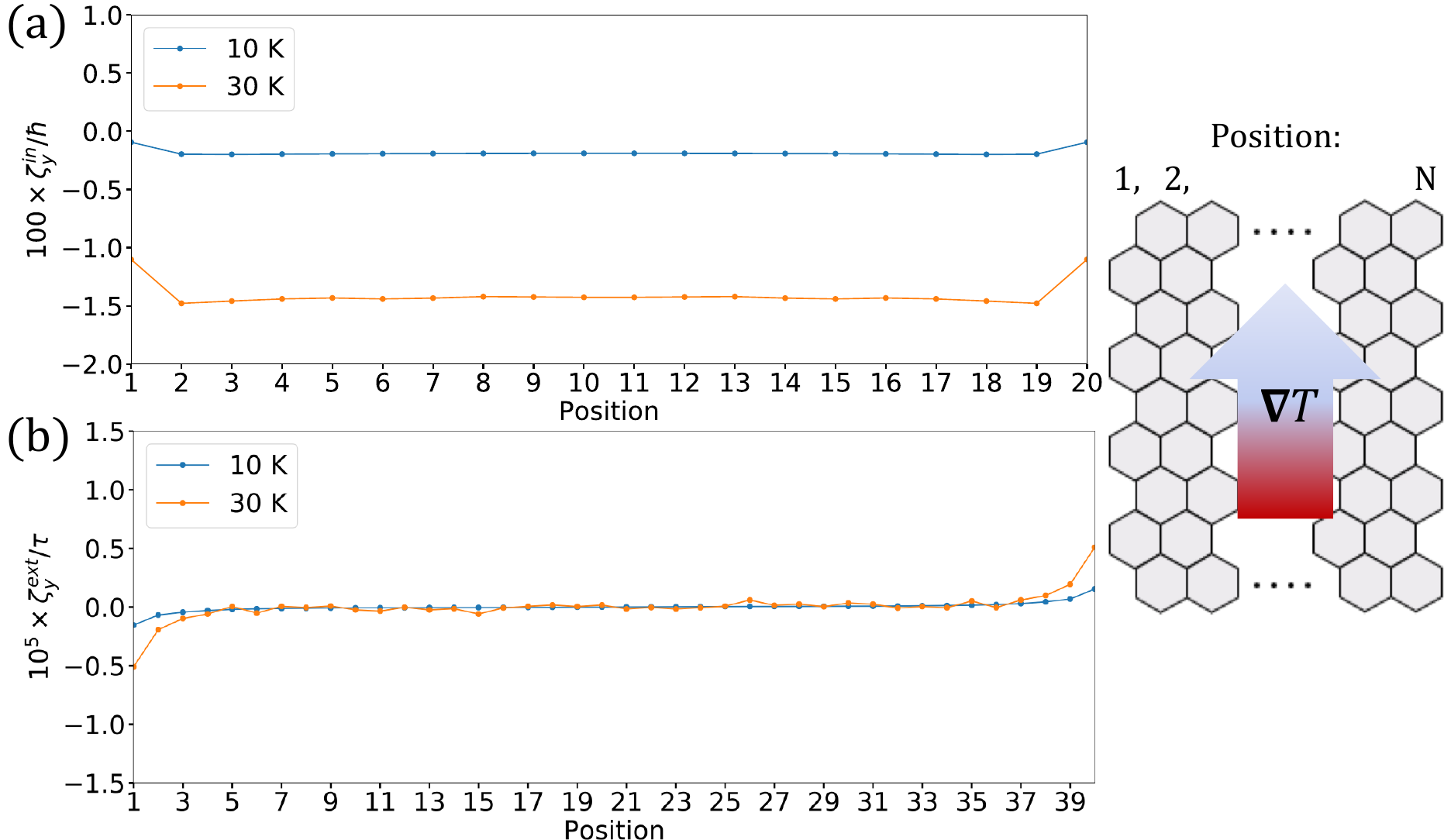}
\caption{(a) $\zeta^{\textrm{in}}_{y}(x)|_{D=0.94}-\zeta^{\textrm{in}}_{y}(x)|_{D=0}$ calculated for $T=10$, $30$ K and with armchair edge as shown on the right. 
(b) $\zeta^{\textrm{ext}}_{y}(x)|_{D=0.94}$  calculated for $T=10$, $30$ K and with armchair edge (note that $\zeta^{\textrm{ext}}_{y}(x)|_{D=0}=0$).
Here, $\tau$ is the lifetime of magnetoelastic excitations.
}
\label{fig.yspin_acc_diff}
\end{figure}

Let us first examine $\zeta_\nu^{\textrm{in}}(r)$ using the Kubo's formula \cite{supplement}.
We find that when the system has zigzag edge, spin density uniformly vanishes whether or not there is MPI.
Similarly, for the armchair edge, we find a symmetric distribution of spin, regardless of the presence of MPI.
On the other hand, $\sum_{x}\zeta^{\textrm{in}}_{y}(x)|_{D\neq 0}\neq 0$ for the armchair edge when there is MPI while $\sum_{x}\zeta^{\textrm{in}}_{y}(x)|_{D=0}=0$ when there is no MPI. 
Thus, although MPI does not induce asymmetric edge spin accumulation through the intrinsic mechanism, it can change the total spin density of the system under thermal gradient.
In Fig.~\ref{fig.yspin_acc_diff} (a), we show the spin density caused by MPI for the armchair edge, i.e. $\zeta^{\textrm{in}}_{y}(x)|_{D\neq 0}-\zeta^{\textrm{in}}_{y}(x)|_{D=0}$.

To observe the spin accumulation induced by spin Nernst current, we need to consider the finite quasiparticle lifetime \cite{murakami2005intrinsic}.  
We study this extrinsic effect using the Boltzmann transport theory within the constant relaxation time approximation \cite{supplement}.
For both the  armchair  and zigzag edges, we find that without MPI, $\zeta_\nu^{\textrm{ext}}(r)|_{D=0}=0$.
However, in the presence of MPI, we find asymmetric distribution of  $\zeta_\nu^{\textrm{ext}}(r)|_{D\neq 0}$ for both the  armchair and zigzag edges.
In particular, the distribution is antisymmetric for the armchair edge, as shown in Fig.~\ref{fig.yspin_acc_diff} (b), so that $\sum_{x}\zeta^{\textrm{ext}}_{y}(x)|_{D\neq 0}=0$.
This is not so for the zigzag edge, and $\sum_{y}\zeta^{\textrm{ext}}_{x}(y)|_{D\neq 0}\neq 0$, so that nonzero total spin density is induced~\cite{supplement}.

In the above, we showed that thermal gradient can induce not only the edge spin accumulation, but also a nonzero total spin density in the system.
For a single layer of ferrimagnet, this will be difficult to detect because not only the temperature gradient but also the change of average temperature can modify the net magnetization.
However, if our model is stacked antiferromagnetically along the $z$ direction such that the neighboring layers are related by a glide mirror $\mathcal{G}_y : (x,y,z)\rightarrow (x,-y,z+\frac{1}{2})$ as in $\textrm{Fe}_2\textrm{Mo}_3\textrm{O}_8$, the spin density induced by thermal gradient in two adjacent layers has the same (opposite) sign for the armchair (zigzag) edge, while the bulk magnetization is always zero. Therefore, the net magnetization for the armchair edge configuration arises just from the thermal gradient, while for the zigzag edge, the net magnetization vanish. 

Temperature gradient can induce nonzero total spin density whenever the spin is not conserved and the system has sufficiently low symmetry, such as broken inversion symmetry. 
The spatial distribution and the total sum of spin also depend strongly on the direction of thermal gradient. 
These behaviors in our model can be explained using symmetry arguments, which is given in the SM.

\textit{Material realization.|}
We suggest that the thermal Hall current arising from MPI may be relevant to $\textrm{Fe}_2\textrm{Mo}_3\textrm{O}_8$.
In Ref.~\cite{ideue2017giant}, giant thermal Hall conductivity of the order $10^{-2}~\textrm{Wm}^{-1}\textrm{K}^{-1}$ was observed in $\textrm{Fe}_2\textrm{Mo}_3\textrm{O}_8$, which was attributed to skew-scattering of phonon. 
According to the phonon scattering mechanism, $\kappa_{xx}\propto \kappa_{xy} \propto \tau_l$, where $\tau_l$ is the phonon lifetime.
However, the data in Ref.~\cite{ideue2017giant} suggests that this relation does not hold at large magnetic field and high temperature. 
We suggest that this may be due to the intrinsic contribution to the thermal Hall conductivity originating from the MPI.
Although the parameters for magnon and phonon are not available to us except for the magnetic moment \cite{varret1972etude}, from Fig.~\ref{fig.conductivity} and the interlayer distance of $\textrm{Fe}_2\textrm{Mo}_3\textrm{O}_8$, we can estimate the order of magnitude of the thermal Hall conductivity arising from the band crossing between one magnon and one phonon bands to be around $0.3\times 10^{-3} \textrm{Wm}^{-1}\textrm{K}^{-1}$ when $D=0.94$ meV. 
Therefore, depending on the material parameters, the MPI can potentially generate a thermal Hall response of order $10^{-3} \textrm{Wm}^{-1}\textrm{K}^{-1}$ to throw off the relation $\kappa_{xx}\propto \kappa_{xy} \propto \tau_l$. 

\textit{Discussion.|} 
In this work, we examined the heat and spin responses arising from MPI in a noncentrosymmetric collinear ferrimagnet with anisotropy.
The unique feature of ferrimagnets is that either of the spin $\pm 1$ magnon band always decreases in energy to hybridize with phonon band when magnetic field is applied parallel to the magnetic order, which is important for thermal Hall and spin Nernst effect.
This is distinct from ferromagnets, in which all of the magnon bands increase or decrease in energy.
Moreover, the intrinsic spin Nernst current originating from MPI induces edge spin accumulation through the extrinsic contribution to the spin density.
This can serve as an indicator of MPI contribution to the thermal Hall conductivity, since magnon does not, by itself, show thermal Hall or spin Nernst effect.
However, since MPI does not conserve the spin, revealing the correspondence between the spin accumulation and the spin current is quite a subtle issue  ~\cite{shi2006proper}, which we leave for future study. 
Finally, we showed that nonzero total spin density induced by thermal gradient can serve as an additional evidence of MPI.

\begingroup
\renewcommand{\addcontentsline}[3]{}
\renewcommand{\section}[2]{}
\begin{acknowledgments}

\textit{Note added|} During the preparation of our manuscript, a related work appeared in which the thermal Hall effect in antiferromagnetic insulators is discussed \cite{zhang2019su3}. 
Although their work also mentions the spin Nernst effect, it is more focused on the $\textrm{SU}(3)$ topology of magnetoelastic excitations and the ensuing thermal Hall conductivity.
We focus more on the correspondence between the spin Nernst current and the spin density.

We thank H. Ishizuka, H. Katsura, and D. Go for helpful discussion. S.P. was supported by IBS-R009-D1. B.-J.Y. was supported by the Institute for Basic Science in Korea (Grant No. IBS-R009-D1) and Basic Science Research Program through the National Research Foundation of Korea (NRF) (Grant No. 0426-20190008), and the POSCO Science Fellowship of POSCO TJ Park Foundation (No. 0426-20180002). This work was supported in part by the U.S. Army Research Office under Grant Number W911NF-18-1-0137. N.N. was supported by JST CREST Grant Number JPMJCR1874, Japan, and JSPS KAKENHI Grant numbers 18H03676 and 26103006.

\end{acknowledgments}
\endgroup
\begingroup
\renewcommand{\addcontentsline}[3]{}
\renewcommand{\section}[2]{}

\endgroup

\newpage
~
\newpage
\appendix

\renewcommand{\appendixpagename}{\center \Large Supplemental Material (SM)}

\appendixpage
\setcounter{page}{1}
\setcounter{figure}{0}
\renewcommand{\appendixname}{SM}
\renewcommand{\thefigure}{{S}\arabic{figure}}
\tableofcontents
\section{Magnon} 

In this section, we study the magnon Hamiltonian given in Eq.~(1) of the main text.
To carry out the spin wave expansion, we introduce the local axes 
\begin{align}
\hat{\bm{x}}_A&=(1,0,0), ~\hat{\bm{y}}_A=(0,1,0),~ \hat{\bm{z}}_A=(0,0,1) \nonumber \\
\hat{\bm{x}}_B&=(1,0,0), ~\hat{\bm{y}}_B=(0,-1,0),~ \hat{\bm{z}}_B=(0,0,-1), \label{eq.local_axes}
\end{align}
so that $\bm{S}_i=S_i^x \hat{\bm{x}}_i+S_i^y \hat{\bm{y}}_i+S_i^y \hat{\bm{z}}_i$ where $i=A,B$. 
To obtain the linear spin wave theory, it suffices to put
\begin{equation}
S_i^x=\frac{\sqrt{2S}}{2}(a_i+a_i^\dagger), ~S_i^y=\frac{\sqrt{2S}}{2i}(a_i-a_i^\dagger),~ S_i^z=S-a_i^\dagger a_i. \label{eq.HP_operators}
\end{equation}
We note any orthonormal system of local axes is valid so long as $\hat{\bm{z}}_i$ points along the magnetic order. 

To the quadratic order in the Holstein-Primakoff (HP) operators, the nearest-neighbor interaction is ($J_1>0$)
\begin{align}
\mathcal{H}_{J_1}&=\sum_{\langle ij \rangle } J_1 \bm{S}_i\cdot \bm{S}_j \nonumber \\
&=-3J_1 N(S^2+S)+\frac{1}{2} \sum_{\bm{k} }\phi_{\bm{k}}^\dagger H_{J_1}(\bm{k}) \phi_{\bm{k}},
\end{align}
where
\begin{equation}
H_{J_1}(\bm{k})=J_1 S\begin{bmatrix}
3 &0&0& \gamma^*_{\bm{k}} \\
0&3&\gamma_{\bm{k}}&0\\
0&\gamma_{\bm{k}}^*&3&0\\
\gamma_{\bm{k}}& 0&0&3
\end{bmatrix}
\end{equation}
and\begin{equation}
\phi_{\bm{k}}=\begin{bmatrix}
a_{A\bm{k}} \\
a_{B \bm{k}} \\
a_{A-\bm{k}}^\dagger \\
a_{B-\bm{k}}^\dagger
\end{bmatrix}.
\end{equation}
We have defined the function $\gamma_{\bm{k}}=\sum_{\bm{\delta}_i}e^{i\bm{k}\cdot \bm{\delta}}$, where $\bm{\delta}_{1}=(0,1)l$, $\bm{\delta}_{2}=(-\frac{\sqrt{3}}{2}, -\frac{1}{2})l$ and $\bm{\delta}_{3}=(\frac{\sqrt{3}}{2},-\frac{1}{2})l$ are the relative positions of the nearest neighbors, and $l$ is the distance between $A$ and $B$.

Similarly, the next-nearest-neighbor Heisenberg interaction is ($J_A,J_B<0$)
\begin{align}
\mathcal{H}_{J_2}&=J_2^A \sum_{\langle \langle ij \rangle \rangle _A}\bm{S}_i\cdot \bm{S}_j+J_2^B \sum_{\langle \langle ij \rangle \rangle _B}\bm{S}_i\cdot \bm{S}_j\nonumber \\
&= 3(J_2^A+J_2^B) N (S^2+S) +  \frac{1}{2}\sum_{\bm{k}}\phi_{\bm{k}}^\dagger H_{J_2}(\bm{k}) \phi_{\bm{k}},
\end{align}
where
\begin{equation}
H_{J_2}(\bm{k}) =S(-6+\tilde{\gamma}_{\bm{k}}) \begin{bmatrix}
J_2^A & 0 & 0 & 0\\
0 & J_2^B & 0 & 0 \\
0 & 0 & J_2^A & 0 \\
0 & 0 & 0 & J_2^B
\end{bmatrix}.
\end{equation}
Here, we defined $\tilde{\gamma}_{\bm{k}}=\sum_{\tilde{\bm{\delta}}} e^{i\bm{k}\cdot \tilde{\bm{\delta}}}$, where $\tilde{\bm{\delta}}$ are the relative positions of the six next-nearest neighbors.

Let us note that the easy axis anisotropy must be along the $z$ direction because of the three-fold rotations about A and B sites. We have ($\alpha_A, \alpha_B<0$)
\begin{align}
\mathcal{H}_{\alpha}&=\sum_{i} \alpha_i(S_i^z)^2 \nonumber \\
&=(\alpha_A+\alpha_B)N(S^2+S)+\frac{1}{2}\sum_{\bm{k}} \phi_{\bm{k}}^\dagger H_{\alpha}(\bm{k}) \phi_{\bm{k}},
\end{align}
where
\begin{equation}
H_{\alpha}(\bm{k})=-2S\begin{bmatrix}
\alpha_A & 0 & 0 & 0\\
0 & \alpha_B & 0 & 0 \\
0 & 0 & \alpha_A & 0 \\
0 & 0 & 0 & \alpha_B
\end{bmatrix}.
\end{equation}

Applying the magnetic field $\bm{H}=H\hat{\bm{z}}$, we have
\begin{align}
\mathcal{H}_H&=\sum_{i}\mu_i \bm{S}_i\cdot \bm{H} \nonumber \\
&=H(\mu_A-\mu_B)N (S+\frac{1}{2})+\frac{1}{2}\sum_{\bm{k}} \phi_{\bm{k}}^\dagger H_H(\bm{k}) \phi_{\bm{k}},
\end{align}
where
\begin{equation}
H_H(\bm{k})=
H\begin{bmatrix}
-\mu_A & 0 & 0 & 0\\
0 & \mu_B & 0 & 0 \\
0 & 0 & -\mu_A & 0 \\
0 & 0 & 0 & \mu_B
\end{bmatrix}.
\end{equation}

Let us briefly discuss the $\mathcal{M}_x \mathcal{C}_{2x}$ symmetry mentioned in the main text.
In the spin sector, both $\mathcal{M}_x$ and $\mathcal{C}_{2x}$ rotates the spin direction about the $x$ axis by $180^\circ$, so that when combined, it does not change the spin direction.
However, $\mathcal{M}_x$ sends the spin at $(x,y)$ to $(-x,y)$.
Thus, the action of the $\mathcal{M}_x \mathcal{C}_{2x}$ symmetry on the HP operators is $a_{i,(k_x,k_y)}\rightarrow a_{i,(-k_x,k_y)}$ for $i=A,B$.
It is easy to see that the magnon Hamiltonian satisfies this symmetry from the expressions above.

\section{Phonon}

In this section, we list the spring constant matrices. 
Using the expression for the spring constant matrix between the nearest neighbor $A$ and $B$, 
\begin{equation}
K(\bm{\delta}_1)=\begin{bmatrix}
-K_T & 0 \\
0 & -K_L
\end{bmatrix},
\end{equation}
and imposing the $\mathcal{C}_{3z}$ symmetry, we have $K(\bm{\delta}_2)=C_{3z}K(\bm{\delta}_1)C_{3z}^T$ and  $K(\bm{\delta}_3)=C_{3z}^TK(\bm{\delta}_1)C_{3z}$, where 
\begin{equation}
C_{3z}=\begin{bmatrix}
-\frac{1}{2}& -\frac{\sqrt{3}}{2} \\
\frac{\sqrt{3}}{2} & -\frac{1}{2}
\end{bmatrix}.
\end{equation}
For the next-nearest neighbors between the $A$ sites, we have
\begin{equation}
K(\bm{\delta}_3-\bm{\delta}_2)=\begin{bmatrix}
-k_L^A & 0 \\
0 & -k_T^A
\end{bmatrix},
\end{equation}
$K(\bm{\delta}_1-\bm{\delta}_3)=C_{3z}K(\bm{\delta}_3-\bm{\delta}_2)C_{3z}^T$ and $K(\bm{\delta}_2-\bm{\delta}_1)=C_{3z}^TK(\bm{\delta}_3-\bm{\delta}_2)C_{3z}$, while the other spring constants follow from the identity $K(\Delta\bm{R})=K(-\Delta \bm{R})$. The spring constant matrices between the $B$ sites is obtained by making the replacement $k_L^A,k_T^A\rightarrow k_L^B, k_T^B$.
Finally, the onsite potentials follow from the condition that there is no change in Hamiltonian from uniform shift of the lattice. 
The contributions from nearest-neighbor interactions to the $A$ and $B$ sites are respectively given by
\begin{align}
K_{nn}(0)&=\frac{1}{2}\begin{bmatrix}
\frac{3}{2}(K_T+K_L) & 0 \\
0 & \frac{3}{2}(K_T+K_L)
\end{bmatrix}, \nonumber \\
 K_{nn}'(0)&=\frac{1}{2}\begin{bmatrix}
\frac{3}{2}(K_T+K_L) & 0 \\
0 & \frac{3}{2}(K_T+K_L)
\end{bmatrix}.
\end{align}
For the next-nearest neighbors, we have
\begin{align}
K_{nnn}(0)=\begin{bmatrix}
\frac{3}{2}(k_L^A+k_T^A) & 0 \\
0 & \frac{3}{2}(k_L^A+k_T^A)
\end{bmatrix}, \nonumber \\
K_{nnn}'(0)=\begin{bmatrix}
\frac{3}{2}(k_L^B+k_T^B) & 0 \\
0 & \frac{3}{2}(k_L^B+k_T^B)
\end{bmatrix}.
\end{align}

\section{Magnon-Phonon Interaction}

As explained in the main text, we consider the magnon-phonon interaction arising from the fluctuation of DM vector direction when the atoms fluctuate from their equilibrium position. To do this, note that when the atoms are at their equilibrium positions, we can write the DM vector as (see Fig.~1 (b) in the main text)  $\bm{D}_{ij}=D\hat{\bm{r}}_{ij}\times \hat{\bm{z}}$, where $\bm{r}_i=\bm{R}_i+\bm{u}_i$ and $\hat{\bm{r}}_{ij}=\frac{\bm{r}_i-\bm{r}_j}{|\bm{r}_i-\bm{r}_j|}$. Assuming for simplicity that $D$ is independent of $\bm{u}_i$ and that $M_A=M_B=M$, the part of $\bm{D}_{ij}$ that is linear in $\bm{u}_{ij}=\bm{u}_i-\bm{u}_j$  is $\left[-\hat{\bm{R}}_{ij}\left(\hat{\bm{R}}_{ij}\cdot\frac{\tilde{\bm{u}}_{ij}}{\sqrt{M}} \right)+\frac{\tilde{\bm{u}}_{ij}}{\sqrt{M}}\right]\times \hat{\bm{z}}\frac{D}{l}$, where $l$ is the distance between the nearest neighboring atoms, $\bm{R}_{ij}=\bm{R}_i-\bm{R}_j$, and $\tilde{\bm{u}}_{ij}=\tilde{\bm{u}}_i-\tilde{\bm{u}}_j$. 
Similarly, the terms linear in the Holstein-Primakoff operators in $\bm{S}_i \times \bm{S}_j$ are given by
$S[S_i^x  \hat{\bm{x}}_i \times \hat{\bm{z}}_j+S_i^y \hat{\bm{y}}_i \times \hat{\bm{z}}_j+S_j^x \hat{\bm{z}}_i \times \hat{\bm{x}}_j+S_j^y \hat{\bm{z}}_i \times \hat{\bm{y}}_j]$, where $\hat{\bm{x}}_i$, $\hat{\bm{y}}_i$, $\hat{\bm{z}}_i$ are the local axes introduced in Eq.~\eqref{eq.local_axes}.
The MPI at the quadratic level is obtained by multiplying these two terms.

Introducing the HP operators as in Eq.~\eqref{eq.HP_operators} and taking the Fourier transformation,  the magnon-phonon interaction is
\begin{align}
&\lambda\sum_{\bm{\delta_i}}\Big\{\big[u_{A\bm{k}}^x(a_{A-\bm{k}}+a_{A\bm{k}}^\dagger+a_{B-\bm{k}}e^{i\bm{k}\cdot \bm{\delta_i}}+a_{B\bm{k}}^\dagger e^{i\bm{k}\cdot \bm{\delta}_i}  )\nonumber \\
&-u_{B\bm{k}}^x (a_{B-\bm{k}}
+a_{B\bm{k}}^\dagger+a_{A-\bm{k}}e^{-i\bm{k}\cdot \bm{\delta_i}}+a_{A\bm{k}}^\dagger e^{-i\bm{k}\cdot \bm{\delta}_i}  ) \nonumber \\
&-iu_{A\bm{k}}^y(a_{A-\bm{k}}-a_{A\bm{k}}^\dagger-a_{B-\bm{k}}e^{i\bm{k}\cdot \bm{\delta_i}}+a_{B\bm{k}}^\dagger e^{i\bm{k}\cdot \bm{\delta}_i}  )\nonumber \\
&-iu_{B\bm{k}}^y(a_{B-\bm{k}}-a_{B\bm{k}}^\dagger-a_{A-\bm{k}}e^{-i\bm{k}\cdot \bm{\delta_i}}+a_{A\bm{k}}^\dagger e^{-i\bm{k}\cdot \bm{\delta}_i}  )
\big]\nonumber \\
&-(\bm{\delta}_i \cdot \hat{\bm{x}})u_{A\bm{k}}^{\delta_i}(a_{A-\bm{k}}+a_{A\bm{k}}^\dagger+a_{B-\bm{k}}e^{i\bm{k}\cdot \bm{\delta_i}}+a_{B\bm{k}}^\dagger e^{i\bm{k}\cdot \bm{\delta}_i}  )\nonumber \\
&+(\bm{\delta}_i \cdot \hat{\bm{x}})u_{B\bm{k}}^{\delta_i}(a_{B-\bm{k}}
+a_{B\bm{k}}^\dagger+a_{A-\bm{k}}e^{-i\bm{k}\cdot \bm{\delta_i}}+a_{A\bm{k}}^\dagger e^{-i\bm{k}\cdot \bm{\delta}_i}  ) \nonumber \\
&+i(\bm{\delta}_i \cdot \hat{\bm{y}})u_{A\bm{k}}^{\delta_i}(a_{A-\bm{k}}-a_{A\bm{k}}^\dagger-a_{B-\bm{k}}e^{i\bm{k}\cdot \bm{\delta_i}}+a_{B\bm{k}}^\dagger e^{i\bm{k}\cdot \bm{\delta}_i}  )\nonumber \\
& +i(\bm{\delta}_i \cdot \hat{\bm{y}})u_{B\bm{k}}^{\delta_i}(a_{B-\bm{k}}-a_{B\bm{k}}^\dagger-a_{A-\bm{k}}e^{-i\bm{k}\cdot \bm{\delta_i}}+a_{A\bm{k}}^\dagger e^{-i\bm{k}\cdot \bm{\delta}_i}  )
\Big\},
\end{align}
where we have defined $u_{A/B}^{\delta_i}\equiv \tilde{\bm{u}}_{A/B}\cdot \bm{\delta}_i$ and $\lambda=\frac{ DS}{2l}\sqrt{\frac{2S}{M}}$

\begin{figure*}[ht]
\centering
\includegraphics[width=17cm]{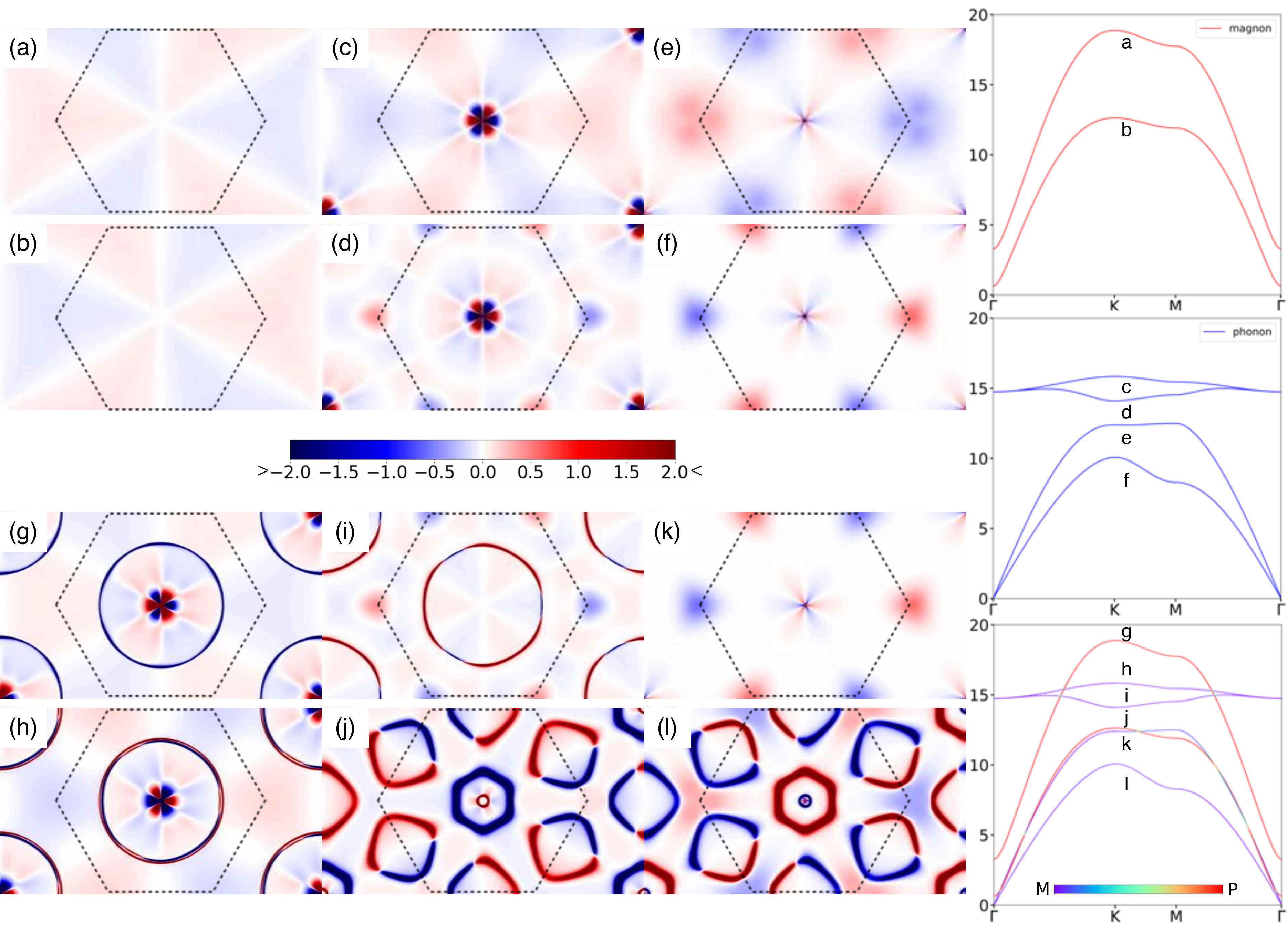}
\caption{ (a) and (b): The Berry curvature density of magnon bands, whose energy spectrums are shown on the right (top) which are labelled accordingly. (c)$\sim$(f): Berry curvature density of phonon bands, whose energy spectrum is shown on the right (middle). (g)$\sim$(l): Berry curvature density of magnetoelastic bands, whose energy spectrum is shown on the right (bottom). The parameters used in the Hamiltonian are the same as in Fig.~(2) (d). Note that the Berry curvature density may go well outside the range of the colorbar. 
}
\label{fig.curvature}
\end{figure*}

\section{ BdG Hamiltonian and Berry curvature}
In this work, we have computed all of the response functions by transforming the magnetoelastic Hamiltonian into bosonic BdG form as explained below. 
Another popular method is to write the Hamiltonian in terms of the phonon operators so that the phonon sector of the Hamiltonian is diagonalized from the outset.
However, this may not result in the correct response functions such as the thermal Hall conductivity, as pointed out in Ref.~\cite{park2019topological}.

To make the connection between the magnetoelastic Hamiltonian and the bosonic BdG Hamiltonian, let us first clarify the relation between the phonon Hamiltonian and bosonic BdG Hamiltonian.
Below, we work in the system of units where $\hbar=1$ and energy is measured in meV.
 If we define 
\begin{equation}
\bm{v}_{A/B\bm{k}}=\frac{\sqrt{2}}{2}(\tilde{\bm{p}}_{A/B\bm{k}}-i\tilde{\bm{u}}_{A/B\bm{k}} ) \label{eq.phonon_bdg_field}
\end{equation}
and
\begin{equation}
\bm{y}_{\bm{k}}=(\bm{v}_{A\bm{k}} , \bm{v}_{B\bm{k}},\bm{v}_{A-\bm{k}}^\dagger,\bm{v}_{B-\bm{k}}^\dagger),
\end{equation}
we have
\begin{equation}
[y_{m \bm{k}},y_{n \bm{k}'}^\dagger ]= (\rho_z)_{mn}\delta_{\bm{k},\bm{k}'}, \quad \bm{y}_{\bm{k}}^\dagger=\rho_x \bm{y}_{-\bm{k}} \label{eq.phonon_bdg_relation}
\end{equation}
where $m$ and $n$ runs over all the components of $\bm{y}_{\bm{k}}$, and $\rho_i$ are the Pauli matrices relating the particle and hole sectors of the field operator $\bm{y}_{\bm{k}}$.
Note that Eq.~\eqref{eq.phonon_bdg_relation} is the defining relation of a bosonic BdG field operator.

We note that in this system of unit where energy is measured in meV and $\hbar=1$, the strength of magnon-phonon interaction is given by $\frac{D}{l} \sqrt{\frac{f \hbar}{M_i}} $ where $f \approx 0.658 \times 10^{-12}$.  
The quantity $f$ is numerically equivalent to $10^3\times \hbar/e$ evaluated in SI units. 
For $l=5\AA$, $M_i=60$ amu, and $D=0.94~\textrm{meV}$, $\frac{D}{l} \sqrt{\frac{f \hbar}{M_i}} \approx 0.050~\textrm{meV/s}^{1/2}$.

Next, let us review the basic properties of BdG Hamiltonian. 
Let $\mathcal{H}=\frac{1}{2} \sum_{\bm{k}}\Psi^\dagger_{\bm{k}}H_{\bm{k}}\Psi_{\bm{k}}$ be the bosonic BdG Hamiltonian where $\Psi_{\bm{k}}$ satisfy \begin{equation}
[\Psi_{m \bm{k}},\Psi_{n \bm{k}'}^\dagger ]= (\tau_z)_{mn}\delta_{\bm{k},\bm{k}'}, \quad \Psi_{\bm{k}}^\dagger=\rho_x \Psi_{-\bm{k}} , \label{eq.bdg_def}
\end{equation}
where $m$, $n=-N,...,-1,1,...,N$.
The transformation to the energy eigenbasis is given by $\Phi_{\bm{k}}=T_{\bm{k}}\Psi_{\bm{k}}$, so that $\mathcal{H}=\frac{1}{2}\sum_{\bm{k}}\Phi_{\bm{k}}^\dagger T_{\bm{k}}^\dagger H_{\bm{k}} T_{\bm{k}}\Phi_{\bm{k}}=\frac{1}{2}\sum_{\bm{k}} \Phi_{\bm{k}}^\dagger E_{\bm{k}}\Phi_{\bm{k}}$, where $E_{\bm{k}}$ is diagonal, and $[\Phi_{\bm{k},m},\Phi^\dagger_{\bm{k}',n}]=(\tau_z)_{m n} \delta_{\bm{k},\bm{k}'}$. 
This requires that $T_{\bm{k}}^\dagger H_{\bm{k}}T_{\bm{k}}=E_{\bm{k}}$ and $T_{\bm{k}}^\dagger \tau_z T_{\bm{k}}=\tau_z$, where $\tau_i$ are the Pauli matrices in the particle and hole sectors. For example,  $(\tau_z)_{mn}=\delta_{mn}$ for $m,n>0$, $(\tau_z)_{mn}=-\delta_{mn}$ for $m,n<0$, and $(\tau_z)_{mn}=0$ otherwise.
Denoting by $|n,\bm{k}\rangle$ the column vectors of $T_{\bm{k}}$ and noticing that $T_{\bm{k}}^{-1}=\tau_z T_{\bm{k}}^\dagger \tau_z$, these conditions translate to $\tau_z H_{\bm{k}}|n,\bm{k}\rangle=E_{\bm{k},n}|n,\bm{k}\rangle$ and $\langle m,\bm{k}| \tau_z |n,\bm{k}\rangle=(\tau_z)_{mn}$.

The Berry curvature for this Hamiltonian is defined to be \cite{matsumoto2014thermal}  $\bm{B}_{\bm{k},n}=i(\tau_z)_{nn}\bm{\nabla}\times \langle n,\bm{k}| \tau_z \bm{\nabla}  | n,\bm{k}\rangle$. In particular, we denote the $z$ component of the Berry curvature by $\Omega_{\bm{k},n}$.
In Fig.~\ref{fig.curvature}, we plot the full Berry curvature density for magnon, phonon, and magnetoelastic bands and the corresponding energy spectrums.
Let us also calculate the Chern numbers, which is defined as $C_n=\frac{1}{2\pi}\int_{BZ} d\bm{k} \Omega_{\bm{k},n}$. 
We find that the individual magnon and phonon bands are topologically trivial. 
Specifically, using the labels for the energy bands in Fig.~\ref{fig.curvature}, we have $C_a=C_b=0$, $C_c+C_d=C_e+C_f=0$.
Note that we have grouped together the bands which are not gapped.
We find that the magnetoelastic bands are topologically nontrivial. 
Specifically, $C_g+C_h=-1$, $C_i=1$, $C_j=C_k+C_l=0$.

\section{Symmetry analysis of thermal Hall and spin Nernst effect}
In the main text, we have stated that  the $\mathcal{M}_x \mathcal{C}_{2x}^S$ symmetry forces $\kappa_{xy}=\alpha^S_{xy}=0$. 
To show this, recall that $\mathcal{M}_x \mathcal{C}_{2x}^S$ is the mirror symmetry about the plane normal to the $x$ axis, which does not change the spin direction. 
It therefore imposes $E_{n,(k_x,k_y)}=E_{n,(-k_x,k_y)}$.
Because Berry curvature transforms like magnetic field in the momentum space, we also have $\Omega_{n,(k_x,k_y)}=-\Omega_{n,(-k_x,k_y)}$.
Using this, the contribution to  $\kappa_{xy}=-\frac{k_B^2 T}{V\hbar }\sum_{\bm{k}}\sum_{n=1}^{N} \left[c_2(g(E_{n,\bm{k}}))-\frac{\pi^2}{3}\right] \Omega_{n,\bm{k}}$ from $(k_x,k_y)$ and $(-k_x,k_y)$ cancel pairwise, so that $\kappa_{xy}=0$.
In addition, because the $\mathcal{M}_x \mathcal{C}_{2x}^S$ symmetry does not change the spin direction, it imposes $\langle \mathcal{S}^z\rangle_{n,(k_x,k_y)}=\langle \mathcal{S}^z \rangle_{n,(-k_x,k_y)}$. 
Using this, the contribution to $\alpha_{xy}^S=-\frac{k_B}{\hbar V}\sum_{\bm{k}}\sum_{n=1}^{N} \langle \mathcal{S}^z \rangle_{n,\bm{k}} \Omega_{n,\bm{k}} c_1(E_{n,\bm{k}}/k_BT)$ from $(k_x,k_y)$ and $(-k_x,k_y)$ cancel pairwise, so that $\alpha^S_{xy}=0$.

\section{Absence of Sign Change in Spin Nernst coefficient}

In this section, we explain why the spin Nernst coefficient computed using the semiclassical theory $\alpha_{xy}^S=-\frac{k_B}{\hbar V}\sum_{\bm{k}}\sum_{n=1}^{N} \langle \mathcal{S}^z \rangle_{n,\bm{k}} \Omega_{n,\bm{k}} c_1(E_{n,\bm{k}}/k_BT)$ does not change sign as temperature is varied, contrary to the case of thermal Hall conductivity.
To this end, it is useful to examine  the momentum space distribution of $\Omega_{n,\bm{k}}^{S}\equiv\frac{1}{\hbar}\langle \mathcal{S}^z \rangle_{n,\bm{k}}\Omega_{n,\bm{k}}=\frac{1}{\hbar}\langle n,\bm{k}| S | n,\bm{k}\rangle \Omega_{n,\bm{k}}$, where the explicit expression for $S^z$ is given in Eq.~\eqref{eq.sz_matrix}. 

\begin{figure}[t]
\centering
\includegraphics[width=8.5cm]{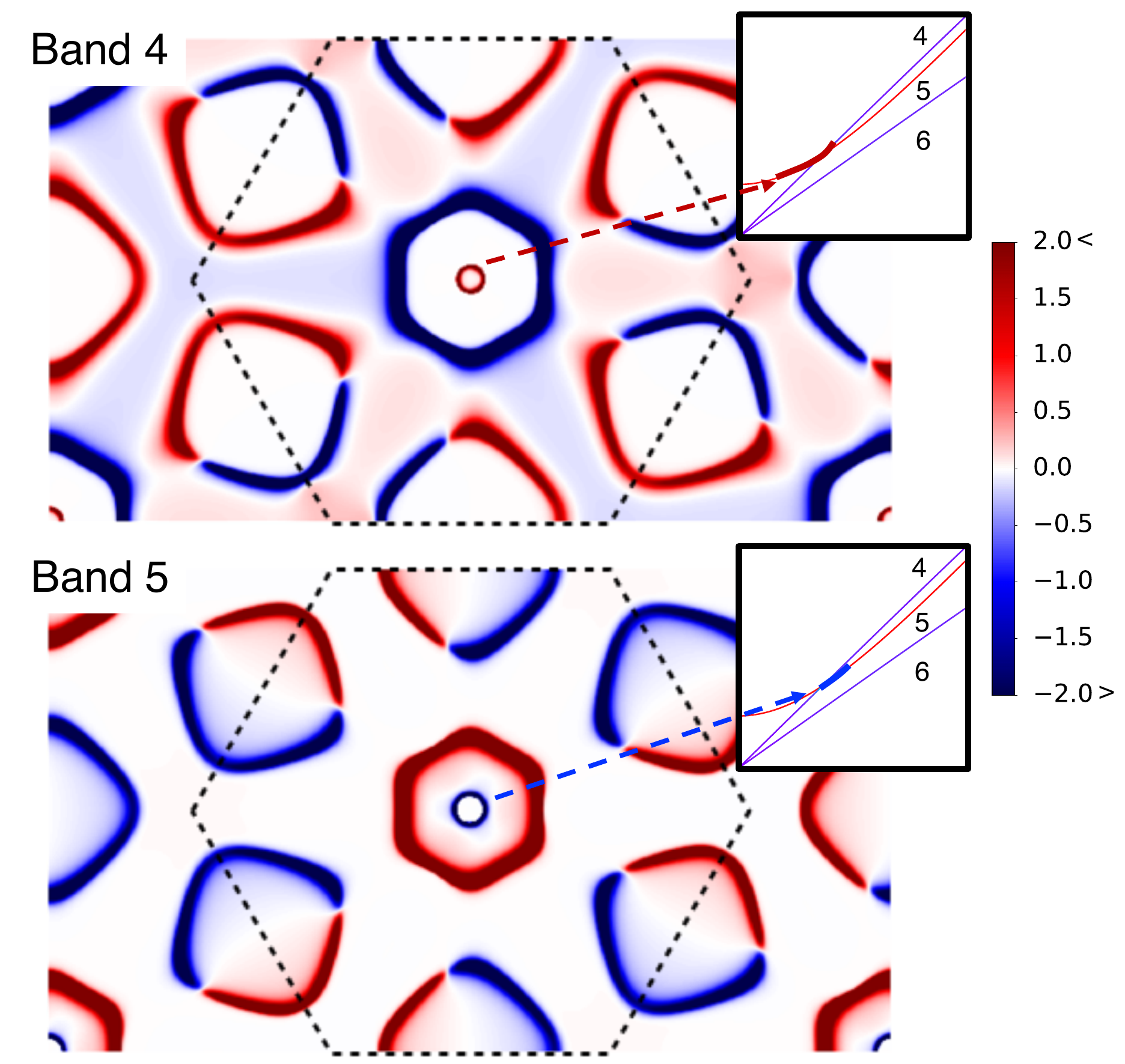}
\caption{ Plot of $\Omega_{n,\bm{k}}^S$ for $n=4$, $5$. The inset shows the energy bands near the $\Gamma$ point, where we have indicated the region relevant to $\alpha^S_{xy}$ at low temperature  with thick red (blue) line when $\Omega_{n,\bm{k}}^S >0$ ($<0$).
}
\label{fig.spin_cvt}
\end{figure}

We plot this quantity in Fig.~\ref{fig.spin_cvt} for $n=4$ and $5$, as they are relevant for the spin Nernst coefficient at low temperatures.
Comparing $\Omega^S_{4,\bm{k}}$ and $\Omega^S_{5,\bm{k}}$, we see that  $\Omega^S_{4,\bm{k}}$ shows a circular region where it takes positive values and has smaller radius. The positive region extends towards the inner regions of the circle and has lower energy.
$\Omega^S_{5,\bm{k}}$, on the other hand, shows a circular region where it takes negative and has larger radius. The negative region extends towards the outer regions of the circle and has higher energy.
Because $c_1(E/k_BT)$ for $E>0$ is positive and decays quickly to $0$ as a function of energy at low temperature, the \textit{positive} contribution to $\alpha^S_{xy}$ from the negative $\Omega^S_{4,\bm{k}}$ in band $4$ around the red circular region is \textit{larger} than the \textit{negative} contribution to $\alpha^S_{xy}$ from the positive $\Omega^S_{5,\bm{k}}$  in band $5$ around the blue circular region at low temperature, which explains why $\alpha_{xy}^S>0$ at low temperature (the temperature scale is set by the energy at which these circular regions occur, which is about $1$ meV). 
Because the region outside of the blue circle in band $5$ quickly becomes red as the radius is increased while the corresponding region in band $4$ stays white, we should expect $\alpha_{xy}^S$ to decrease at larger temperature, which is consistent with the numerical calculation in Fig.~(3) (b) in the main text. 

\section{Spin Nernst Effect}
Much of the linear response theory used here was already developed in Refs.~\cite{zyuzin2016magnon,matsumoto2014thermal}, but we repeat them for completeness.
Let
\begin{equation}
\Psi_{\bm{k}}=\left[a_{A\bm{k}},~a_{B\bm{k}}, ~\bm{v}_{\bm{k}},~ a_{A-\bm{k}}^\dagger, ~a_{B-\bm{k}}^\dagger,~ \bm{v}_{-\bm{k}}^\dagger\right]
\end{equation}
be the bosonic BdG basis, where $a_{A/B\bm{k}}$ are the Holstein-Primakoff operators and $\bm{v}_{A/B\bm {k}}$ are the bosonic BdG fields for phonons defined in Eq.~\eqref{eq.phonon_bdg_field}. 
Then, 
\begin{equation}
\mathcal{H}_{0}=\frac{1}{2}\sum_{\bm{k}}\Psi^{\dagger}_{\bm{k}} H_{\bm{k}} \Psi_{\bm{k}}, \quad \mathcal{S}^z=\frac{1}{2} \sum_{\bm{k}}\Psi^\dagger_{\bm{k}} S^z \Psi_{\bm{k}}, \label{eq.H_k}
\end{equation}
where $\mathcal{H}_{0}$ is the magnetoelastic Hamiltonian and $\mathcal{S}^z$ is the spin \textit{excitation} operator, which should not be confused with $S_i^z$, which was defined as the spin operator along the direction of the magnetic order in Eq.~\eqref{eq.HP_operators}. 
In matrix form, we have
\begin{equation}
S^z=\hbar \begin{bmatrix}
-1 & & & & & \\
 & 1 & & & & \\
 & & 0_{4} & & & \\
 & & &  -1 & & \\
 & & & & 1 & \\
 & & & & & 0_{4}
\end{bmatrix}, \label{eq.sz_matrix}
\end{equation}
where $0_4$ is the $4\times4$ zero-matrix.
We find
\begin{align}
[\mathcal{H}_{0},\mathcal{S}^z]=\frac{1}{2}\sum_{\bm{k}} \Psi^\dagger_{\bm{k}} (H_{\bm{k}} \tau_z S^z-S^z\tau_z H_{\bm{k}} )\Psi_{\bm{k}}.
\end{align}
Explicitly evaluating $H_{\bm{k}} \tau_z S^z-S^z\tau_z H_{\bm{k}}$, one finds that only terms that couple magnon and phonon survives. 
Namely, $\mathcal{H}_0$ and $\mathcal{S}^z$ commute when the magnon-phonon interaction is neglected.

While the spin Nernst effect in Refs.~\cite{zyuzin2016magnon,cheng2016spin} was analyzed for the case when spin is conserved, here we extend the theory to the case when the spin is not conserved.

\subsection{Method of pseudo-gravitational potential}
Let us study the spin current for a general BdG Hamiltonian in response to thermal gradient.
Below, repeated Roman indices implies summation over the BdG field operator indices taking the values  $-N, -N-1, ... -1, 1, ..., N-1, N$. 
Calligraphic letters are reserved for operators containing BdG fields.
Finally, we put $\hbar=1$ and restore it at the end.

Let 
\begin{align}
\mathcal{H}_0&=\frac{1}{2} \sum_{\bm{\delta}}\int d\bm{r} \Psi_m^\dagger(\bm{r}) H_{mn}^{\bm{\delta}}\Psi_n(\bm{r}+\bm{\delta}) \nonumber \\
&= \frac{1}{2} \sum_{\bm{\delta}} \int d\bm{r} \Psi^\dagger (\bm{r}) H^{\bm{\delta}} e^{i \bm{p} \cdot \bm{\delta}} \Psi(\bm{r})\nonumber \\
&\equiv \frac{1}{2}\int d\bm{r} \Psi^\dagger(\bm{r}) h_0 \Psi(\bm{r}) \label{eq.magnon_h_0} \nonumber \\
&\equiv\frac{1}{2}\int d\bm{r} \mathcal{h}_0(\bm{r}) .
\end{align}
Here, $\bm{p}$ is the momentum operator.
Taking the Fourier transformation of the field operators
\begin{equation}
\Psi(\bm{r})=\frac{1}{\sqrt{V}}\sum_{\bm{k}} \Psi_{\bm{k}}e^{i\bm{k}\cdot \bm{r}},
\end{equation}
we find that
\begin{equation}
H_{\bm{k}}=\sum_{\bm{\delta}} e^{i\bm{k}\cdot \bm{\delta}} H^{\bm{\delta}},
\end{equation}
where $H_{\bm{k}}$ was defined in Eq.~\eqref{eq.H_k}.

According to Luttinger \cite{luttinger1964theory}, one can compute the response to thermal gradient by introducing the gravitational scalar potential $\chi (\bm{r})$ that couples to the Hamiltonian. 
Assuming that the potential is linear in position, i.e. $\chi(\bm{r})=\bm{r}\cdot \bm{\nabla}\chi$, this interaction is given by
\begin{equation}
{\cal V}=\frac{1}{4}\int d\bm{r} \Psi^\dagger(\bm{r}) (h_0 \bm{r}+\bm{r} h_0) \Psi(\bm{r}) \cdot \bm{\nabla} \chi.
\end{equation}
The total Hamiltonian,
\begin{equation}
\mathcal{H}=\mathcal{H}_0+\mathcal{V},
\end{equation}
is equivalent up to linear order in $\chi(\bm{r})$ to
\begin{equation}
\mathcal{H}=\frac{1}{2} \sum_{\bm{\delta}}\int d\bm{r} \tilde{\Psi}_m^\dagger(\bm{r}) H_{mn}^{\bm{\delta}}\tilde{\Psi}_n(\bm{r}+\bm{\delta}).
\end{equation}
where
\begin{equation}
\tilde{\Psi}_n(\bm{r})=(1+\frac{\bm{r}\cdot \bm{\nabla}\chi}{2})\Psi_n(\bm{r})\equiv \xi(\bm{r}) \Psi_n(\bm{r}). \label{eq.tilde_psi}
\end{equation}
Because of the relation \cite{luttinger1964theory}
\begin{equation}
\langle j^S_{\mu}\rangle=L^S_{\mu \nu}\left(T\nabla_\nu \frac{1}{T}-\nabla_{\nu}\chi \right)
\end{equation}
we have $\alpha_{\mu \nu}^S=L^S_{\mu \nu}/T$.
\subsection{Current operator}
Let
\begin{equation}
{\cal S}^z(\bm{r})=\frac{1}{2}\Psi_m^\dagger(\bm{r}) S^z_{mn}\Psi_n(\bm{r}),
\end{equation}
be the spin excitation operator. Its time evolution is given by
\begin{widetext}
\begin{align}
-i\frac{\partial {\cal S}(\bm{r})}{\partial t}=&[\mathcal{H}, {\cal S}(\bm{r})]\nonumber \\
=&\frac{1}{4} \sum_{\bm{\delta}}\int d\bm{r}' \left([\tilde{\Psi}^\dagger_m(\bm{r}') H^{\bm{\delta}}_{mn}\tilde{\Psi}_n(\bm{r}'+\bm{\delta}),\Psi^\dagger_k(\bm{r})]S^z_{kl}(\bm{r}) \Psi_l(\bm{r})-\Psi_k^\dagger(\bm{r})S^z_{kl}[\Psi_l(\bm{r}),\tilde{\Psi}^\dagger_m(\bm{r}')H^{\bm{\delta}}_{mn}\tilde{\Psi}_{n}(\bm{r}'+\bm{\delta})]\right) \nonumber \\
=&\frac{1}{4}\sum_{\bm{\delta}} \left((-i\tau_y)_{mk} \xi(\bm{r})H_{mn}^{\bm{\delta}} \xi(\bm{r}+\bm{\delta})\Psi_n(\bm{r}+\bm{\delta})
+\Psi^{\dagger}_m(\bm{r}-\bm{\delta})\xi(\bm{r}-\bm{\delta})H_{mn}^{\bm{\delta}}\xi(\bm{r})(\tau_z)_{nk} \right)S^z_{kl}\Psi_l(\bm{r}) \nonumber \\
&-\frac{1}{4}\sum_{\bm{\delta}} \Psi^\dagger_k(\bm{r}) S^z_{kl} \left((\tau_z)_{lm}\xi(\bm{r})H_{mn}^{\bm{\delta}}\xi(\bm{r}+\bm{\delta})\Psi_n (\bm{r}+\bm{\delta})
+\Psi_m^\dagger(\bm{r}-\bm{\delta})\xi(\bm{r}-\bm{\delta})H_{mn}^{\bm{\delta}} \xi(\bm{r})(i\tau_y)_{ln}   \right), \label{eq.current_intermediate}
\end{align}
where we have used
\begin{equation}
[\Psi_{m}(\bm{r}),\Psi_{n}^\dagger(\bm{r}')]=(\tau_z)_{mn} \delta(\bm{r}-\bm{r}'), \quad [\Psi^\dagger_{m}(\bm{r}),\Psi^\dagger_{n}(\bm{r}')]=-(i\tau_y)_{mn} \delta(\bm{r}-\bm{r}'), \quad [\Psi_{m}(\bm{r}),\Psi_{n}(\bm{r}')]=i(\tau_y)_{mn} \delta(\bm{r}-\bm{r}') .\label{eq.magnon_commutation}
\end{equation}
The third line of Eq.~\eqref{eq.current_intermediate} containing $\tau_y$ can be manipulated by using 
\begin{equation}
\Psi^\dagger_m (\bm{r})=(\tau_x)_{mn} \Psi_n  (\bm{r}), \quad \tau_x S^z \tau_x =(S^z)^T, \quad \tau_x H^{\bm{\delta}}\tau_x=(H^{-\bm{\delta}})^T  .\label{eq.magnon_properties}
\end{equation}
Note that the first equality follows from definition of BdG field. The second and the third follow from the first equality.  
The third line of Eq.~\eqref{eq.current_intermediate} becomes
\begin{align}
\frac{1}{4}\sum_{\bm{\delta}} \tilde{\Psi}^{\dagger}(\bm{r}+\bm{\delta})H^{-\bm{\delta}} \tau_z S^z \tilde{\Psi}(\bm{r})-\tilde{\Psi}^{\dagger}(\bm{r}) S^z \tau_z H^{\bm{-\delta}} \tilde{\Psi} (\bm{r}-\bm{\delta}).
\end{align}
In total, we obtain
\begin{align}
&\frac{1}{2}\sum_{\bm{\delta}} \Psi^{\dagger}(\bm{r}-\bm{\delta})\xi(\bm{r}-\bm{\delta})H^{\bm{\delta}}\xi(\bm{r})\tau_zS^z\Psi(\bm{r})
-\Psi^\dagger(\bm{r}) S^z\tau_z\xi(\bm{r})H^{\bm{\delta}}\xi(\bm{r}+\bm{\delta})\Psi (\bm{r}+\bm{\delta}).
\end{align}
Thus,
\begin{align}
-i\frac{\partial \mathcal{S}^z (\bm{r})}{\partial t}=\frac{1}{2}\sum_{\bm{\delta}} \tilde{\Psi}^{\dagger}(\bm{r}-\bm{\delta})H^{\bm{\delta}} \tau_zS^z\tilde{\Psi}(\bm{r})
-\tilde{\Psi}^\dagger(\bm{r}) S^z\tau_z H^{\bm{\delta}} \tilde{\Psi} (\bm{r}+\bm{\delta}).
\end{align}
Now, note that 
\begin{align}
\bm{v}=i[\mathcal{H},\bm{r}]=i\sum_{\bm{\delta}}\bm{\delta}H^{\bm{\delta}}e^{i\bm{p}\cdot \bm{\delta}}. \label{eq.magnon_velocity}
\end{align}
Thus,
\begin{align}
\frac{\partial \mathcal{S}^z (\bm{r})}{\partial t}=&-\frac{i}{2}\sum_{\bm{\delta}}\big(\tilde{\Psi}^\dagger(\bm{r})S^z\tau_z H^{\bm{\delta}} \tilde{\Psi}(\bm{r}+\bm{\delta})-\tilde{\Psi}^\dagger(\bm{r}-\bm{\delta})S^z\tau_z H^{\bm{\delta}}\tau_z S^z \
\tilde{\Psi}(\bm{r})+\tilde{\Psi}^\dagger(\bm{r}-\bm{\delta})S^z\tau_z H^{\bm{\delta}}\tilde{\Psi}(\bm{r}) \nonumber \\
&~~~~~~~~~~~~~~-\tilde{\Psi}^\dagger(\bm{r}-\bm{\delta})H^{\bm{\delta}}\tau_z S^z \tilde{\Psi}(\bm{r})+\tilde{\Psi}^\dagger(\bm{r})H^{\bm{\delta}}\tau_z S^z \tilde{\Psi}(\bm{r}+\bm{\delta})-\tilde{\Psi}^\dagger(\bm{r})H^{\bm{\delta}}\tau_z S^z \tilde{\Psi}(\bm{r}+\bm{\delta})   \big) \nonumber \\
&=-\frac{i}{2} \sum_{\bm{\delta}} \bigg(\bm{\nabla} \cdot \big(\tilde{\Psi}^\dagger(\bm{r}) S^z \tau_z \bm{\delta}H^{\bm{\delta}} \tilde{\Psi}(\bm{r}+\bm{\delta})+ \tilde{\Psi}^\dagger (\bm{r}) \bm{\delta}H^{\bm{\delta}} \tau_z S^z \tilde{\Psi}(\bm{r}+\bm{\delta})\big) \nonumber \\
&~~~~~~~~~~~~~~~~~~~~~~+\tilde{\Psi}^\dagger(\bm{r}-\bm{\delta})S^z\tau_z H^{\bm{\delta}} \tilde{\Psi}(\bm{r})-\tilde{\Psi}^\dagger(\bm{r}-\bm{\delta}) H^{\bm{\delta}} \tau_z S^z \tilde{\Psi}(\bm{r})\bigg) \nonumber \\
&=-\bm{\nabla}\cdot \tilde{\Psi}^\dagger(\bm{r})\frac{S^z\tau_z \bm{v}+\bm{v}\tau_z S^z}{2}\tilde{\Psi}(\bm{r})-\frac{i}{2}\sum_{\bm{\delta}}\tilde{\Psi}^\dagger(\bm{r})(S^z\tau_z H^{\bm{\delta}}e^{i\bm{p}\cdot \bm{\delta}}-H^{\bm{\delta}}e^{i\bm{p}\cdot \bm{\delta}}\tau_z S^z)\tilde{\Psi}(\bm{r}) .
\end{align}
Therefore, we define the spin current operator as
\begin{equation}
\bm{j}_{S}=\tilde{\Psi}^\dagger(\bm{r}) \frac{\bm{v}\tau_z S^z + S^z \tau_z \bm{v}}{2}\tilde{\Psi}(\bm{r}).
\end{equation}

\subsection{Linear response to temperature gradient}
To linear order in temperature gradient, we have
\begin{align}
\bm{j}_{S}(\bm{r})&=\frac{1}{2}\Psi^{\dagger}(\bm{r})(S^z \tau_z \bm{v}+\bm{v}\tau_z S^z)\Psi(\bm{r})+\frac{\nabla_ \mu \chi}{4}(S^z \tau_z \bm{v} r_ \mu+ S^z\tau_z r_ \mu  \bm{v}+ \bm{v}   r_ \mu \tau_z S^z+r_ \mu \bm{v} \tau_z S^z ) \nonumber \\
&=\bm{j}_S^{(0)}+\bm{j}_S^{(1)},
\end{align}
respectively.
Define $\bm{J}_S\equiv \int d\bm{r} \bm{j}_S(\bm{r})$. To linear order in temperature gradient, 
\begin{equation}
\langle \bm{J}_S\rangle=\langle \bm{J}_S^{(0)} \rangle_{\textrm{neq}}+\langle \bm{J}_S^{(1)} \rangle_{\textrm{eq}}.
\end{equation}
Notice that $\bm{J}_S^{(0)}$ should be averaged over the non-equilibrium distribution, while $\bm{J}_S^{(1)}$, which is already linear in the temperature gradient, should be averaged over the equilibrium distribution.

From now on, we drop the subscript $S$ on the spin current.

\subsection{Evaluation of  Kubo formula } \label{sec.evaluation_kubo}

The Kubo formula is
\begin{equation}
\langle J^{(0)}_\mu \rangle_{\textrm{neq}}=- \lim_{\omega\rightarrow 0} \frac{\partial}{\partial \omega} \int_{0}^{\beta} d\tau e^{i\omega \tau} \langle T_{\tau} J^{(0)}_{\mu}(\tau) J^{Q}_{\nu}(0) \rangle  \nabla_\nu \chi\equiv - S_{\mu \nu} \nabla_\nu \chi,
\end{equation}
where $\omega=2\pi n/\beta$ and $n$ is an integer. 
Here, $J^{Q}_\mu$ satisfies $\frac{\partial \cal{V}}{\partial t}= \bm{J}^{Q}\cdot \bm{\nabla} \chi$. 
Since
\begin{align}
\frac{\partial {\cal V}}{\partial t}&=i[\mathcal{H}_0,{\cal V}] \nonumber \\
&=\frac{i}{4}\bm{\nabla} \chi \cdot \sum_{\bm{\delta},\bm{\delta}'}\int d\bm{r} \Psi^\dagger(\bm{r}+\bm{\delta}')H^{-\bm{\delta}'}(-\tau_z)(2\bm{r}+\bm{\delta}') H^{\bm{\delta}} \Psi(\bm{r}+\bm{\delta})+\Psi^\dagger(\bm{r})H^{\bm{\delta}} \tau_z H^{\bm{\delta}'}(2\bm{r}+\bm{\delta}+\bm{\delta}') \Psi(\bm{r}+\bm{\delta}+\bm{\delta}') \nonumber \\
&=\frac{i}{4}\bm{\nabla}\cdot \chi \sum_{\bm{\delta},\bm{\delta}'}\int d\bm{r} \Psi^\dagger(\bm{r}) H^{\bm{\delta}} \tau_z (2\bm{r}+2\bm{\delta}+\bm{\delta}') H^{\bm{\delta}'}\Psi(\bm{r}+\bm{\delta}+\bm{\delta}')-\Psi^\dagger(\bm{r}) H^{\bm{\delta}} \tau_z (2\bm{r}+\bm{\delta}) H^{\bm{\delta}'} \Psi(\bm{r}+\bm{\delta}+\bm{\delta}') \nonumber \\
&=\frac{i}{4}\bm{\nabla}\cdot \chi \sum_{\bm{\delta},\bm{\delta}'}\int d\bm{r} \Psi^\dagger(\bm{r})H^{\bm{\delta}}\tau_z (\bm{\delta}+\bm{\delta}')H^{\bm{\delta}'}\Psi(\bm{r}+\bm{\delta}+\bm{\delta}') \nonumber \\
&=\bm{J}^Q \cdot \bm{\nabla} \chi ,
\end{align}
we have
\begin{equation}
\bm{J}^Q=\frac{1}{4}\int d\bm{r} \Psi^\dagger(\bm{r}) (h_0 \tau_z \bm{v}+\bm{v} \tau_z h_0 )\Psi(\bm{r}). \label{eq.magnon_heat_current}
\end{equation}
To obtain the second line, we used Eqs.~\eqref{eq.magnon_commutation} and \eqref{eq.magnon_properties}, and to obtain the third line, we shifted the integration variable.
Finally, to obtain Eq.~\eqref{eq.magnon_heat_current}, we use Eqs.~\eqref{eq.magnon_h_0} and \eqref{eq.magnon_velocity}.

Taking the Fourier transform, we have
\begin{align}
\bm{J}^{(0)}&=\frac{1}{2} \sum_{\bm{k},\bm{\delta}}\Psi^\dagger_{\bm{k}}[S^z \tau_z \bm{\delta} H^{\bm{\delta}}e^{i\bm{k}\cdot \bm{\delta}}+\bm{\delta} H^{\bm{\delta}}e^{i\bm{k}\cdot \bm{\delta}} \tau_z S^z ]\Psi_{\bm{k}} \nonumber \\
&\equiv \frac{1}{2}\sum_{\bm{k}} \Psi^\dagger_{\bm{k}}[S^z \tau_z \bm{v}_{\bm{k}}+ \bm{v}_{\bm{k}}\tau_z S^z]\Psi_{\bm{k}}
\end{align}
and
\begin{align}
\bm{J}^{Q}&=\frac{1}{4}\sum_{\bm{k},\bm{\delta},\bm{\delta}'} \Psi^\dagger_{\bm{k}}[H^{\bm{\delta}}e^{i\bm{k}\cdot \bm{\delta}} \tau_z \bm{\delta}' H^{\bm{\delta}'}e^{i\bm{k}\cdot\bm{\delta}'}+\bm{\delta}H^{\bm{\delta}}e^{i\bm{k}\cdot \bm{\delta}} \tau_z  H^{\bm{\delta}'}e^{i\bm{k}\cdot\bm{\delta}'}]\Psi_{\bm{k}}\nonumber \\
&\equiv \frac{1}{4}\sum_{\bm{k}}\Psi^\dagger_{{\bm{k}}}[H_{\bm{k}}\tau_z \bm{v}_{\bm{k}}+\bm{v}_{\bm{k}}\tau_z H_{\bm{k}}   ]\Psi_{\bm{k}}.
\end{align}
Let us introduce the field operators for the energy eigenstates (see below Eq.~\eqref{eq.bdg_def}),
\begin{equation}
\Phi_{\bm{k}}=T_{\bm{k}} \Psi_{\bm{k}}.
\end{equation}
We then have
\begin{align}
S_{\mu \nu}=\frac{1}{8}\lim_{\omega\rightarrow 0}\frac{\partial}{\partial \omega} \int_{0}^{\beta} e^{i\omega \tau} \sum_{\bm{k},\bm{k}'}& \langle\Phi^\dagger_{\bm{k}}(\tau) T^\dagger_{\bm{k}}(S^z \tau_z v_{\bm{k},\mu}+v_{\bm{k},\mu} \tau_z S^z)T_{\bm{k}}\Phi_{\bm{k}}(\tau) \nonumber \\
&\times\Phi^\dagger_{\bm{k}'}(0) T^\dagger_{\bm{k}'}(H_{\bm{k}'}\tau_z v_{\bm{k}',\nu}+v_{\bm{k}',\nu}\tau_z H_{\bm{k}'}    ) T_{\bm{k}'}\Phi_{\bm{k}'}(0)\rangle.
\end{align}
Note the identity 
\begin{align}
\langle \Phi^\dagger_{\bm{k},m}(\tau) \Phi_{\bm{k},n}(\tau) \Phi^\dagger_{\bm{k}',p}(0)\Phi_{\bm{k}',q}(0)\rangle=&\langle \Phi^\dagger_{\bm{k},m} (\tau)\Phi_{\bm{k},n}(\tau) \rangle \langle \Phi^\dagger_{\bm{k}',p} (0)\Phi_{\bm{k}',q}(0) \rangle
+\langle \Phi^\dagger_{\bm{k},m} (\tau)\Phi^\dagger_{\bm{k}',p}(0) \rangle\langle \Phi_{\bm{k},n} (\tau)\Phi_{\bm{k}',q}(0) \rangle \nonumber \\
&+\langle \Phi^\dagger_{\bm{k},m} (\tau)\Phi_{\bm{k}',q}(0) \rangle\langle \Phi_{\bm{k},n} (\tau)\Phi^\dagger_{\bm{k}',p}(0) \rangle.
\end{align}
The integral $\int_{0}^{\beta} e^{i\omega \tau}\langle \Phi^\dagger_{\bm{k}} (\tau)\Phi_{\bm{k}}(\tau) \rangle \langle \Phi^\dagger_{\bm{k}'} (0)\Phi_{\bm{k}'}(0) \rangle$ vanishes (because $\omega=2\pi n/\beta$), so we only need the correlation functions between different times.
We have
\begin{align}
\langle \Phi^\dagger_{\bm{k},m}(\tau) \Phi^\dagger_{\bm{k}',n}(0)\rangle&=\delta_{\bm{k},-\bm{k}'}(i\tau_y)_{mn}g[(\tau_z E_{\bm{k}})_{mm}]e^{(\tau_z E_{\bm{k}})_{mm}\tau} \\
\langle \Phi_{\bm{k},m}(\tau) \Phi_{\bm{k}',n}(0)\rangle&=\delta_{\bm{k},-\bm{k}'}(-i\tau_y)_{mn}g[-(\tau_z E_{\bm{k}})_{mm}]e^{-(\tau_z E_{\bm{k}})_{mm}\tau}\\
\langle \Phi^\dagger_{\bm{k},m}(\tau) \Phi_{\bm{k}',n}(0)\rangle&=\delta_{\bm{k},\bm{k}'}(\tau_z)_{mn}g[(\tau_z E_{\bm{k}})_{mm}]e^{(\tau_z E_{\bm{k}})_{mm}\tau}\\
\langle \Phi_{\bm{k},m}(\tau) \Phi^\dagger_{\bm{k}',n}(0)\rangle&=\delta_{\bm{k},\bm{k}'}(-\tau_z)_{mn}g[-(\tau_z E_{\bm{k}})_{mm}]e^{-(\tau_z E_{\bm{k}})_{mm}\tau},
\end{align}
where $g[x]=\frac{1}{e^{\beta x}-1}$ is the Bose-Einstein distribution.
If we define $\bm{V}^S_{\bm{k}}=T^{\dagger}_{\bm{k}}[S^z \tau_z \bm{v}_{\bm{k}} +\bm{v}_{\bm{k}}\tau_zS^z]T_{\bm{k}}$ and $\bm{V}_{\bm{k}}=T^\dagger_{\bm{k}} \bm{v}_{\bm{k}}T_{\bm{k}}$, we have
\begin{align}
S_{\mu \nu}=\frac{1}{8}\lim_{\omega\rightarrow 0}\frac{\partial}{\partial \omega} \int_{0}^{\beta} e^{i\omega \tau} \sum_{\bm{k},\bm{k}'} &[V^S_{\bm{k},\mu}]_{mn}[E_{\bm{k}'}\tau_z V_{\bm{k}',\nu}+V_{\bm{k}',\nu} \tau_z E_{\bm{k}'}]_{pq}  [(\tau_y)_{mp} (\tau_y)_{nq}\delta_{\bm{k},-\bm{k}'} -(\tau_z)_{mq} (\tau_z)_{np}\delta_{\bm{k},\bm{k}'}]\nonumber \\
&\times g[(\tau_z E_{\bm{k}})_{mm}]g[-(\tau_z E_{\bm{k}})_{nn}]e^{[(\tau_z E_{\bm{k}})_{mm}-(\tau_z E_{\bm{k}})_{nn}]\tau}.
\end{align}
The integral is
\begin{equation}
\lim_{\omega\rightarrow 0} \frac{\partial}{\partial \omega} \int_0^\beta e^{(i\omega+(\tau_z E_{\bm{k}})_{mm}-(\tau_z E_{\bm{k}})_{nn})\tau}=\lim_{\omega\rightarrow 0}\frac{\partial}{\partial \omega}\frac{e^{\beta(\tau_z E_{\bm{k}})_{mm}-\beta(\tau_z E_{\bm{k}})_{nn} }-1}{i\omega+(\tau_z E_{\bm{k}})_{mm}-(\tau_z E_{\bm{k}})_{nn}}
=-i\frac{e^{\beta(\tau_z E_{\bm{k}})_{mm}-\beta(\tau_z E_{\bm{k}})_{nn} }-1}{[(\tau_z E_{\bm{k}})_{mm}-(\tau_z E_{\bm{k}})_{nn}]^2}.
\end{equation}
Using the identity
\begin{equation}
[g(x)-g(y)]=-g(x)g(-y)[e^{\beta x-\beta y}-1],
\end{equation}
we have
\begin{equation}
S_{\mu \nu}=\frac{i}{8}\sum_{\bm{k},\bm{k}'} [V^S_{\bm{k},\mu}]_{mn}[E_{\bm{k}'}\tau_z V_{\bm{k}',\nu}+V_{\bm{k}',\nu} \tau_z E_{\bm{k}'}]_{pq} [(\tau_y)_{mp} (\tau_y)_{nq}\delta_{\bm{k},-\bm{k}'} -(\tau_z)_{mq} (\tau_z)_{np}\delta_{\bm{k},\bm{k}'} ]\frac{g[(\tau_z E_{\bm{k}})_{mm}]-g[(\tau_z E_{\bm{k}})_{nn}]}{[(\tau_z E_{\bm{k}_{mm}})-(\tau_z E_{\bm{k}_{nn}})]^2}.
\end{equation}
Now, note the identities
\begin{equation}
E_{-\bm{k}}=\tau_x E_{\bm{k}} \tau_x, \quad T_{-\bm{k}}=\tau_x T_{\bm{k}}^* \tau_x, \quad \bm{v}_{-\bm{k}}=-\tau_x \bm{v}_{\bm{k}}^T \tau_x.
\end{equation}
The first two follows from $H_{-\bm{k}}=\tau_x H_{\bm{k}}^T \tau_x$. 
More generally, $T_{-\bm{k}}=(T_{\bm{k}}P_{\bm{k}})^*$, where $P_{\bm{k}}$ is such that $E_{\bm{-k}}=\tau_x P^\dagger_{\bm{k}} E_{\bm{k}} P_{\bm{k}} \tau_x$.
However, the results do not depend on the gauge choice, so for simplicity, we put  $P_{\bm{k}}=1$ \cite{zyuzin2016magnon,matsumoto2014thermal}.
Finally, the  third identity is obtained as follows:
\begin{align}
\bm{v}_{-\bm{k}}= \sum_{\bm{\delta}} \bm{\delta} H^{\bm{\delta}} e^{-i\bm{k}\cdot \bm{\delta}}=\sum_{\bm{\delta}} \tau_x \bm{\delta} H^{-\bm{\delta}} \tau_x e^{-i\bm{k}\cdot \bm{\delta}}=-\sum_{\bm{\delta}} \tau_x \bm{\delta} H^{\bm{\delta} e^{i\bm{k}\cdot \bm{\delta}}} \tau_x=-\tau_x\bm{v}_{\bm{k}}^T \tau_x.
\end{align}
Using these identities, we can manipulate the term containing $\tau_y$ as follows:
\begin{align}
&\frac{i}{8}\sum_{\bm{k},\bm{k}'} [V^S_{\bm{k},\mu}]_{mn}[E_{\bm{k}'}\tau_z V_{\bm{k}',\nu}+V_{\bm{k}',\nu} \tau_z E_{\bm{k}'}]_{pq} (\tau_y)_{mp} (\tau_y)_{nq}\delta_{\bm{k},-\bm{k}'} \frac{g[(\tau_z E_{\bm{k}})_{mm}]-g[(\tau_z E_{\bm{k}})_{nn}]}{[(\tau_z E_{\bm{k}_{mm}})-(\tau_z E_{\bm{k}_{nn}})]^2} \nonumber \\
=&\frac{i}{8}\sum_{\bm{k}} [V^S_{\bm{k},\mu}]_{mn}[(\tau_y)(E_{-\bm{k}}\tau_z T_{-\bm{k}}^\dagger v_{-\bm{k},\nu} T_{-\bm{k}}+T_{-\bm{k}}^\dagger v_{-\bm{k},\nu} T_{-\bm{k}} \tau_z E_{-\bm{k}})  (-\tau_y)]_{mn}\frac{g[(\tau_z E_{\bm{k}})_{mm}]-g[(\tau_z E_{\bm{k}})_{nn}]}{[(\tau_z E_{\bm{k}_{mm}})-(\tau_z E_{\bm{k}_{nn}})]^2} \nonumber \\
=&\frac{i}{8}\sum_{\bm{k}} [V^S_{\bm{k},\mu}]_{mn}[(\tau_y)(\tau_x E_{\bm{k}} \tau_z ( T_{\bm{k}}^T v^{T}_{\bm{k},\nu} T_{\bm{k}}^* \tau_x)+(\tau_x T_{\bm{k}}^T v^{T}_{\bm{k},\nu} T_{\bm{k}}^* )\tau_z E_{\bm{k}}\tau_x)  (-\tau_y)]_{mn}\frac{g[(\tau_z E_{\bm{k}})_{mm}]-g[(\tau_z E_{\bm{k}})_{nn}]}{[(\tau_z E_{\bm{k}_{mm}})-(\tau_z E_{\bm{k}_{nn}})]^2} \nonumber \\
=&-\frac{i}{8}\sum_{\bm{k}} [V^S_{\bm{k},\mu}]_{mn}[\tau_z (T_{\bm{k}}^\dagger v_{\bm{k},\nu} T_{\bm{k}} \tau_z E_{\bm{k}} + E_{\bm{k}} \tau_{z}T_{\bm{k}}^\dagger v_{\bm{k},\nu}T_{\bm{k}})\tau_z]_{nm}\frac{g[(\tau_z E_{\bm{k}})_{mm}]-g[(\tau_z E_{\bm{k}})_{nn}]}{[(\tau_z E_{\bm{k}_{mm}})-(\tau_z E_{\bm{k}_{nn}})]^2} .
\end{align}
Therefore, the term containing $\tau_y$ is identical to the term containing $\tau_z$.
Thus,
\begin{align}
S_{\mu \nu}&=-\frac{i}{4}\sum_{\bm{k}} [V^S_{\bm{k},\mu}]_{mn}[\tau_z (T_{\bm{k}}^\dagger v_{\bm{k},\nu} T_{\bm{k}} \tau_z E_{\bm{k}} + E \tau_zT_{\bm{k}}^\dagger v_{\bm{k},\nu}T_{\bm{k}})\tau_z]_{nm}\frac{g[(\tau_z E_{\bm{k}})_{mm}]-g[(\tau_z E_{\bm{k}})_{nn}]}{[(\tau_z E_{\bm{k}_{mm}})-(\tau_z E_{\bm{k}_{nn}})]^2} .
\end{align}
If we define
\begin{equation}
\bm{w}_{\bm{k}}=S^z \tau_z \bm{v}_{\bm{k}}, \quad \bm{u}_{\bm{k}}=\bm{v}_{\bm{k}} \tau_z S^z ,
\label{eq.w_and_u}
\end{equation}
this is equal to
\begin{align}
S_{\mu \nu} =-\frac{i}{4} \sum_{\bm{k}}  [\langle n | w_{\mu} |m \rangle+\langle n | u_{\mu} | m\rangle] [E_{nn} \langle m |v_\nu |n\rangle (\tau_z)_{mm}+E_{mm} \langle m | v_\nu | n\rangle (\tau_z)_{nn}]\frac{g[(\tau_z E)_{nn}]-g[(\tau_z E)_{mm}]}{[(\tau_z E)_{nn}-(\tau_z E)_{mm}]^2},
\end{align}
where we omit $\bm{k}$ dependence.

\subsection{Method of Smrcka and Streda }

Our goal here is to evaluate 
\begin{equation}
\langle J_{\mu}^{(1)}\rangle_{\textrm{eq}}=-M_{\mu \nu} \nabla_\nu \chi
\end{equation}
First, we note that
\begin{align}
\bm{J}^{(1)} =\int d\bm{r} \Psi^\dagger (\bm{r}) \textrm{Sym}[S^z \tau_z \bm{v} \chi(\bm{r})] \Psi(\bm{r}) = \sum_{\bm{k}}\Psi_{\bm{k}}^\dagger \textrm{Sym}[S^z \tau_z \bm{v}_{\bm{k}} \chi(\bm{r})]\Psi_{\bm{k}}=\sum_{\bm{k}} \Phi_{\bm{k}}^\dagger T_{\bm{k}} \textrm{Sym}[S^z \tau_z \bm{v} \chi(\bm{r})] T_{\bm{k}} \Phi_{\bm{k}},
\end{align}
where $\textrm{Sym}[S^z \tau_z \bm{v} \chi(\bm{r})]=\frac{1}{4}\{ S^z \tau_z ,\{\bm{v} ,\chi(\bm{r}) \}\}$ is the symmetrization. Note that the $\bm{r}$ in $\chi (\bm{r})$ becomes $i\frac{\partial}{\partial \bm{k}}$. 
Taking the expectation value, we obtain
\begin{equation}
\langle \bm{J}^{(1)}\rangle_{\textrm{eq}}=\sum_{\bm{k}} \textrm{Tr}\left[ \tau_z T_{\bm{k}}^\dagger \textrm{Sym} [S^z \tau_z \bm{v}_{\bm{k}} \chi(\bm{r})]T_{\bm{k}} g[\tau_z E_{\bm{k}}]\right].
\end{equation}
Using the identity
\begin{equation}
e^{\tau_z E_{\bm{k}}} =\tau_z T^\dagger_{\bm{k}} \tau_z e^{\tau_z H_{\bm{k}}} T_{\bm{k}},
\end{equation}
we have
\begin{equation}
\langle J_\mu^{(1)}\rangle_{\textrm{eq}}=\sum_{\bm{k}} \nabla_\nu \chi\textrm{Tr}\left[ \tau_z  \textrm{Sym} [S^z \tau_z v_{\bm{k},\mu} r_{\nu}] g[\tau_z H_{\bm{k}}]\right] = M_{\mu \nu}\nabla_\nu \chi.
\end{equation}
Recalling the definition in Eq.~\eqref{eq.w_and_u}, we have
\begin{equation}
M_{\mu \nu}=-\frac{1}{4}\sum_{\bm{k}} \int d\eta g(\eta) \textrm{Tr} [\tau_z (w_{\bm{k},\mu} r_\nu +r_\nu w_{\bm{k},\mu}+u_{\bm{k},\mu} r_\nu+r_\nu u_{\bm{k},\mu})\delta(\eta-\tau_z H_{\bm{k}})].
\end{equation}
Below, we will often omit the $\bm{k}$ dependence for notational simplicity.

Define
\begin{align}
A_{\mu \nu}&=\frac{i}{2} \textrm{Tr}\left[ \tau_z w_\mu \frac{dG^{+}}{d\eta}\tau_z v_\nu \delta(\eta-\tau_z H)-\tau_z w_\mu \delta(\eta-\tau_z H)\tau_z v_\nu \frac{dG^{-}}{d\eta} \right] \\
B_{\mu \nu}&=\frac{i}{2}\textrm{Tr} \left[ \tau_z w_\mu G^+ \tau_z v_\nu \delta(\eta-\tau_z H)-\tau_z w_\mu \delta(\eta-\tau_z H) \tau_z v_\nu G^{-}\right],
\end{align}
where
\begin{equation}
G^{\pm}=\frac{1}{\eta\pm i\epsilon-\tau_z H}
\end{equation}
satisfies
\begin{equation}
i\delta(\eta-\tau_z H)=-\frac{1}{2\pi} (G^+-G^-), \quad \frac{dG^{\pm}}{d\eta}=-(G^{\pm})^2, \quad i \frac{d}{d\eta} \delta(\eta-\tau_z H)=\frac{1}{2\pi}((G^+)^2-(G^-)^2) .\label{eq.green_identities}
\end{equation}
Using these, we find
\begin{align}
A_{\mu \nu}-\frac{1}{2} \frac{dB_{\mu \nu}}{d\eta}&=\frac{1}{8\pi}\textrm{Tr} \left[ -\tau_z w_\mu \frac{dG^+}{d\eta} \tau_z v_\nu G^+
-\tau_z w_\mu G^- \tau_z v_\nu \frac{dG^-}{d\eta}
+\tau_z w_\mu G^+ \tau_z v_\nu \frac{dG^+}{d\eta}
+\tau_z w_\mu  \frac{dG^-}{d\eta}  \tau_z v_\nu G^-\right] \nonumber \\
&=\frac{1}{8\pi}\textrm{Tr} \left[ \tau_z w_\mu (G^+)^2 \tau_z v_\nu G^+
+\tau_z w_\mu G^- \tau_z v_\nu (G^-)^2
-\tau_z w_\mu G^+ \tau_z v_\nu (G^+)^2
-\tau_z w_\mu (G^-)^2 \tau_z v_\nu G^-  \right] .
\end{align}
Note that
\begin{equation}
\bm{v}=i[\bm{r},\tau_z (G^\pm)^{-1}] , \quad \bm{w}=iS^z \tau_z [\bm{r},\tau_z (G^\pm)^{-1}] , \quad \bm{u}=i[\bm{r},\tau_z (G^\pm)^{-1}] \tau_z S^z.\label{eq.u_and_w}
\end{equation}
Using the first identity, we obtain
\begin{align}
A_{\mu \nu}-\frac{1}{2} \frac{dB_{\mu \nu}}{d\eta}=\frac{i}{8\pi}\textrm{Tr} \left[
\tau_z w_\mu [(G^+)^2-(G^-)^2] r_\nu+\tau_z w_\mu r_\nu [(G^+)^2-(G^-)^2]+2\tau_z w_\mu [G^- r_\nu G^- -G^+ r_\nu G^+] \right].
\end{align}
The last term causes problem because of the non-commutation of the Hamiltonian and the spin.
Defining $C_S=\tau_z [H,S^z \tau_z]$, the second term becomes 
\begin{equation}
-\frac{1}{4\pi} \textrm{Tr} [S^z \tau_z r_\mu r_\nu (G^--G^+) -S^z \tau_z r_\mu (G^--G^+)r_\nu+C_S r_\mu G^+ r_\nu G^+ - C_S r_\mu G^- r_\nu G^-].
\end{equation}
Because $[r_\mu,r_\nu]=0$, the first two terms cancel. Using Eq.~\eqref{eq.green_identities}, we obtain
\begin{align}
A_{\mu \nu}-\frac{1}{2} \frac{dB_{\mu \nu}}{d\eta}=&-\frac{1}{4}\textrm{Tr}[\tau_z w_\mu \frac{d}{d\eta} \delta(\eta-\tau_z H) r_\nu+\tau_z w_\mu r_\nu \frac{d}{d\eta} \delta(\eta-\tau_z H)] \nonumber \\
&+\frac{i}{2}\textrm{Tr}[C_S r_\mu  \delta(\eta-\tau_z H) r_\nu G^++C_S r_\mu G^- r_\nu  \delta(\eta-\tau_z H)].
\end{align}
Let us call the first line $m_{\mu \nu}^{(0)}$ and the second line $m_{\mu \nu}^{(1)}$. Replacing $\bm{w}$ by $\bm{u}$, we have
\begin{align}
\tilde{A}_{\mu \nu}-\frac{1}{2} \frac{d\tilde{B}_{\mu \nu}}{d\eta}=&-\frac{1}{4}\textrm{Tr}[\tau_z u_\mu \frac{d}{d\eta} \delta(\eta-\tau_z H) r_\nu+\tau_z u_\mu  r_\nu \frac{d}{d\eta} \delta(\eta-\tau_z H)] \nonumber \\
&+\frac{i}{2}\textrm{Tr}[r_\mu C_S  \delta(\eta-\tau_z H) r_\nu G^++r_\mu  C_S G^- r_\nu  \delta(\eta-\tau_z H)].
\end{align}
Let us call the first line $\tilde{m}_{\mu \nu}^{(0)}$ and the second line $\tilde{m}_{\mu \nu}^{(1)}$.
We may further simplify the terms containing $C_S$. 
Using the trace formula and the completeness relation  
\begin{equation}
\textrm{Tr}(A)=\sum_{n} (\tau_z)_{nn}\langle n| \tau_z A |n\rangle, \quad 1=\sum_{n} | n\rangle (\tau_z)_{nn}\langle n|\tau_z \label{eq.completeness},
\end{equation}
we have
\begin{align}
\frac{i}{2}\textrm{Tr}[C_S r_\mu \delta(\eta-\tau_z H) r_\nu G^+ ] &= \frac{i}{2}  (\tau_z)_{nn} \langle n|\tau_z C_S r_\mu |m \rangle (\tau_z)_{mm} \langle m | \tau_z \delta(\eta-\tau_z H) r_\nu G^+ |n\rangle \nonumber \\
&=\frac{i}{2}(\tau_z)_{nn}\langle n | \tau_z C_S r_\mu |m\rangle (\tau_z)_{mm} \delta(\eta-\tau_z E)_{mm} \langle m | \tau_z r_\nu |n\rangle \frac{1}{(\tau_z E)_{mm}-(\tau_z E)_{nn}+i\epsilon},
\end{align}
and similarly,
\begin{align}
\frac{i}{2}\textrm{Tr} [C_S r_\mu G^- r_\nu \delta(\eta-\tau_z H)]&= \frac{i}{2} (\tau_z)_{nn} \langle n | \tau_z C_S r_\mu G^- |m\rangle (\tau_z)_{mm}\langle m|\tau_z r_\nu \delta(\eta - \tau_z H) |n\rangle \nonumber \\
&=\frac{i}{2}(\tau_z)_{nn}\langle n|\tau_z C_S r_\mu |m\rangle \frac{1}{(\tau_z E)_{nn}-(\tau_z E)_{mm}-i\epsilon} (\tau_z)_{mm}\langle m|\tau_z r_\nu |n\rangle \delta(\eta  - \tau_z E)_{nn}.
\end{align}
Note that we have used $\tau_z H|n\rangle=(\tau_z E)_{nn}|n\rangle$ and its Hermitian conjugate. Note that the terms with $n=m$ cancel.
Thus, 
\begin{align}
m^{(1)}_{\mu \nu}=\sum_{n\neq m}\frac{i}{2}(\tau_z)_{nn} (\tau_z)_{mm} \langle n | \tau_z C_S r_\mu | m\rangle \langle m | \tau_z r_\nu | n\rangle \left[\frac{\delta(\eta-\tau_z E)_{mm}}{(\tau_z E)_{mm}-(\tau_z E)_{nn}}+\frac{\delta(\eta-\tau_z E)_{nn}}{(\tau_z E)_{nn}-(\tau_z E)_{mm}}\right],
\end{align}
and similarly
\begin{align}
\tilde{m}^{(1)}_{\mu \nu}=\sum_{n\neq m}\frac{i}{2}(\tau_z)_{nn} (\tau_z)_{mm} \langle n | \tau_z r_\mu   C_S| m\rangle \langle m | \tau_z r_\nu | n\rangle \left[\frac{\delta(\eta-\tau_z E)_{mm}}{(\tau_z E)_{mm}-(\tau_z E)_{nn}}+\frac{\delta(\eta-\tau_z E)_{nn}}{(\tau_z E)_{nn}-(\tau_z E)_{mm}}\right].
\end{align}
Note that $(\tilde{m}^{(1)}_{\mu \nu})^*=m^{(1)}_{\mu \nu}$ because $C_S^\dagger=-\tau_z C_S \tau_z$.
Using these expressions, we can see that for a bounded spectrum, 
\begin{equation}
\int_{-\infty}^{\infty} d\eta \left[ A_{\mu \nu}(\eta) -\frac{d B_{\mu \nu}(\eta)}{d\eta}\right]=0, \quad \int_{-\infty}^{\infty} d\eta \left[ \tilde{A}_{\mu \nu}(\eta) -\frac{d \tilde{B}_{\mu \nu}(\eta)}{d\eta}\right]=0 \label{eq.AB_sum_rule}.
\end{equation}

Thus,
\begin{align}
M_{\mu \nu}=&-\sum_{\bm{k} }\int_{-\infty}^{\infty} d\eta g(\eta) \int_{\eta}^{\infty}d\tilde{\eta} \left[A_{\mu \nu}(\tilde{\eta}) -\frac{1}{2} \frac{\partial B_{\mu \nu}(\tilde{\eta})}{\partial \tilde{\eta}}+\tilde{A}_{\mu \nu} (\tilde{\eta})-\frac{1}{2} \frac{\partial \tilde{B}_{\mu \nu}(\tilde{\eta})}{\partial \tilde{\eta}} -m^{(1)}_{\mu \nu}-\tilde{m}^{(1)}_{\mu \nu}\right] .
\end{align}
Let us call the first four terms $M^{(0)}_{\mu \nu}$ and last two terms $M^{(1)}_{\mu \nu}$. Using Eq.~\eqref{eq.AB_sum_rule}, we have
\begin{align}
M^{(0)}_{\mu \nu}=&-\sum_{\bm{k} }\left(\int_{-\infty}^{\infty} d\tilde{\eta} \int_{\eta}^\infty (-g(\eta)) \right)\left[A_{\mu \nu}(\tilde{\eta}) -\frac{1}{2} \frac{\partial B_{\mu \nu}(\tilde{\eta})}{\partial \tilde{\eta}}+\tilde{A}_{\mu \nu} (\tilde{\eta})-\frac{1}{2} \frac{\partial \tilde{B}_{\mu \nu}(\tilde{\eta})}{\partial \tilde{\eta}} \right]\nonumber \\
=&-\sum_{\bm{k} }\int_{-\infty}^\infty d\tilde{\eta} \left[A_{\mu \nu}(\tilde{\eta}) -\frac{1}{2} \frac{\partial B_{\mu \nu}(\tilde{\eta})}{\partial \tilde{\eta}}+\tilde{A}_{\mu \nu} (\tilde{\eta})-\frac{1}{2} \frac{\partial \tilde{B}_{\mu \nu}(\tilde{\eta})}{\partial \tilde{\eta}} \right] \int_{\tilde{\eta}}^{\infty} d\eta (-g(\eta)) \label{eq.M0}
\end{align}
Similarly,
\begin{align}
M^{(1)}_{\mu \nu}=\sum_{\bm{k}}\textrm{Re} \left[ i\sum_{m\neq n} \frac{(\tau_z)_{nn} (\tau_z)_{mm} \langle n | \tau_z    r_\mu C_S| m \rangle \langle m|\tau_z r_\nu | n\rangle } {(\tau_zE)_{mm}-(\tau_zE)_{nn}} \int_{(\tau_z E)_{mm}}^{(\tau_z E)_{nn} } d\eta (-g(\eta)) \right]
\end{align}

To evaluate Eq.~\eqref{eq.M0}, it is convenient to express $A_{\mu \nu}-\frac{\partial B_{\mu \nu}}{\partial \eta}$ in terms of Green's function and delta functions 
\begin{align}
A_{\mu \nu}-\frac{1}{2} \frac{\partial B_{\mu \nu}}{\partial \eta}=&\frac{i}{4} \textrm{Tr} \bigg[ \tau_z w_\mu \frac{dG^+}{d\eta}\tau_z v_\nu \delta(\eta-\tau_z H) -\tau_z v_\nu \frac{dG^-}{d\eta} \tau_z w_\mu \delta(\eta -\tau_z H)  \nonumber \\
& ~~~~~~~~~+\tau_z v_\nu G^- \tau_z w_\mu \frac{d}{d\eta} \delta(\eta-\tau_z H) -\tau_z w_\mu G^+ \tau_z v_\nu \frac{d}{d\eta} \delta(\eta-\tau_z H)\bigg].
\end{align}
Using Eq.~\eqref{eq.completeness}, we find that its contribution to $M^{(0)}_{\mu \nu}$  is 
\begin{align}
M^{(0)}_{\mu \nu} =&-\frac{i}{2} \sum_{\bm{k}} \int_{-\infty}^{\infty} d\eta \int_{\eta}^{\infty} (-g(\tilde{\eta}))d\tilde{\eta} \bigg[ (\tau_z)_{nn}\langle n | v_\nu | m\rangle \frac{(\tau_z)_{mm}}{(\eta-\tau_z E_{mm}-i\epsilon)^2}\langle m | w_\mu | n\rangle \delta(\eta-(\tau_z E)_{nn}) \nonumber \\
&\;\;\;\;\;\;\;\;\;\;\;\;\;\;\;\;\;\;\;\;\;\;\;\;\;\;\;\;\;\;\;\;\;\;\;\;\;\; -(\tau_z)_{nn} \langle n | \omega_\mu | m\rangle \frac{(\tau_z)_{mm}}{(\eta-\tau_z E_{mm}+i\epsilon)^2} \langle m | v_\nu | n \rangle \delta(\eta-(\tau_z E)_{nn})  \bigg] \nonumber \\
&-\frac{i}{4} \sum_{\bm{k}}\int_{-\infty}^{\infty} d\eta\bigg[ (\tau_z)_{nn} \langle n | w_\mu | m\rangle \frac{(\tau_z)_{mm}}{\eta-(\tau_z E)_{mm}+i\epsilon} \langle m | v_\nu | n\rangle g(\eta) \delta(\eta-(\tau_z E)_{nn}) \nonumber \\ 
& \;\;\;\;\;\;\;\;\;\;\;\;\;\;\;\;\;\;\;\; -(\tau_z)_{nn} \langle n | v_\nu | m\rangle \frac{(\tau_z)_{mm}}{\eta-(\tau_z E)_{mm}-i\epsilon} \langle m | w_\mu | n\rangle g(\eta) \delta(\eta-(\tau_z E)_{nn})\bigg]\nonumber \\
=&-\frac{i}{2}\sum_{\bm{k},m\neq n}\langle n | w_\mu | m \rangle \langle m | v_\nu | n\rangle \frac{(\tau_z)_{mm}(\tau_z)_{nn}}{((\tau_z E_{mm})-(\tau_z E_{nn}))^2} \left( \int_{(\tau_z E)_{nn}}^{\infty}d\eta g(\eta)-\int_{(\tau_z E)_{mm}}^{\infty}d\eta g(\eta)\right) \nonumber \\
&-\frac{i}{4} \sum_{\bm{k},m\neq n} (\tau_z)_{nn}(\tau_z)_{mm} \langle n | w_\mu | m \rangle \langle m | v_\nu | n \rangle  \frac{g(\tau_z E_{nn})+g(\tau_z E_{mm})}{(\tau_z E)_{nn}-(\tau_z E)_{mm}} .
\end{align}
To obtain the first equality, we cyclicly permute the delta functions to the right and then integrate by parts. 
Thus, 
\begin{align}
S_{\mu \nu}+M^{(0)}_{\mu \nu}=&-\frac{i}{2} \sum_{\bm{k},m\neq n} \langle n | w_\mu+u_\mu | n\rangle \langle m|v_\nu | n\rangle  \frac{(\tau_z)_{mm} E_{nn} g(\tau_z E_{nn})-g(\tau_z E_{mm})E_{mm} (\tau_z)_{nn}  }{((\tau_z E)_{nn}-(\tau_z E)_{mm})^2} \nonumber \\
&-\frac{i}{2}\sum_{\bm{k},m\neq n}\langle n | w_\mu +u_\mu| m \rangle \langle m | v_\nu | n\rangle \frac{(\tau_z)_{mm}(\tau_z)_{nn}}{((\tau_z E_{mm})-(\tau_z E_{nn}))^2} \left( \int_{(\tau_z E)_{nn}}^{\infty}d\eta g(\eta)-\int_{(\tau_z E)_{mm}}^{\infty}d\eta g(\eta)\right) \nonumber \\
=&-\frac{i}{2}\sum_{\bm{k},m\neq n} \langle n | w_\mu + u_\mu |m\rangle \langle m |v_\nu | n\rangle \frac{(\tau_z)_{mm} (\tau_z)_{nn}}{((\tau_z E)_{mm}-(\tau_z E)_{nn})^2} \int_{(\tau_z E)_{mm}}^{(\tau_z E)_{nn}} d\eta \eta \frac{d g(\eta)}{d\eta }.
\end{align}
Changing the variable to $\eta=k_B T x$ and restoring the $\hbar$, the spin Nernst conductivity $\alpha^S_{\mu \nu}=\frac{S_{\mu \nu}+M_{\mu \nu}}{V T}$ is given by
\begin{align}
\alpha^S_{\mu \nu}=&-k_B \hbar \frac{1}{V} \frac{i}{2}\sum_{\bm{k},m\neq n} \langle n | w_\mu + u_\mu |m\rangle \langle m |v_\nu | n\rangle \frac{(\tau_z)_{mm} (\tau_z)_{nn}}{((\tau_z E)_{mm}-(\tau_z E)_{nn})^2} \int_{(\tau_z E)_{mm}/k_BT}^{(\tau_z E)_{nn}/k_BT} dx x \frac{d \rho(x)}{dx } \nonumber \\
&-\frac{k_B}{\hbar} \frac{1}{V} \sum_{\bm{k},m\neq n}\textrm{Re} \left[ i\frac{(\tau_z)_{nn} (\tau_z)_{mm} \langle n | \tau_z r_\mu C_S | m \rangle \langle m|\tau_z r_\nu | n\rangle } {(\tau_zE)_{mm}-(\tau_zE)_{nn}} \int_{(\tau_z E)_{mm}/k_BT}^{(\tau_z E)_{nn}/k_BT } dx \rho(x) \right],
\end{align}
where $\rho(x)=\frac{1}{e^{x}-1}$.
The second term is difficult to evaluate because we need to find (locally) differentiable wave function throughout the Brillouin zone. As explained in more detail below, we expect this term to be small in comparison to the first because it is proportional to $C_S$ which is proportional to the magnon-phonon interaction strength, which is about $0.05$ for $D=0.94$ (in units where we measure energy in meV and $\hbar=1$). 

Now, a note on the integrals. Using
\begin{equation}
\rho(-x)=-1-\rho(x),
\end{equation}
we have
\begin{equation}
\int_0^{-E} \rho(x)dx=\int_0^{E} (1+\rho(x)) dx, \quad \int_0^{-E} x \frac{d\rho(x)}{dx} dx=\int_0^{E} x \frac{d\rho(x)}{dx}.
\end{equation}
It follows that
\begin{align}
\int_{a}^{b} dx \rho(x)=\tilde{c}(b)-\tilde{c}(a),  \quad \int_{a}^{b} dx x \frac{dg (x)}{dx}=c_1(|b|)-c_1(|a|) ,
\end{align}
where $\tilde{c}(x)=\log \left| 1-e^{-|x|}\right|+|x| \theta(-x)$ and  $c_1(x)=\int_{x}^\infty dx' x' \left(-\frac{d\rho(x')}{dx'}\right)=(1+\rho(x))\log(1+\rho(x))-\rho(x) \log \rho(x)$.

We also have
\begin{align}
v_\mu=\frac{1}{\hbar} \frac{\partial H}{\partial k_\mu}, \quad w_\mu=\frac{1}{\hbar}  S^z \tau_z \frac{\partial H}{\partial k_\mu}, \quad u_\mu=\frac{1}{\hbar} \frac{\partial H}{\partial k_\mu}  \tau_z S^z,
\end{align}
so that
\begin{align}
\alpha_{\mu \nu}^{S}\approx \frac{k_B}{\hbar} \frac{i}{2V}\sum_{\bm{k},m\neq n} \left\langle n \left| S^z \tau_z \frac{\partial H}{\partial k_\mu} + \frac{\partial H}{\partial k_\mu} \tau_z S^z \right| m \right\rangle & \left\langle m \left| \frac{\partial H}{\partial k_\nu} \right| n\right\rangle  \frac{(\tau_z)_{mm}(\tau_z)_{nn}}{((\tau_z E)_{mm}-(\tau_z E)_{nn})^2}\nonumber \\ 
& \times \left[c_1 \left(\frac{E_{mm}}{k_BT}\right)-c_1\left(\frac{E_{nn}}{k_BT}\right)\right], \label{eq.approx_spin_nernst}
\end{align}
where we have neglected the term containing $C_S$.

To justify this, let us note that
\begin{align}
\left\langle n \left| \frac{\partial H}{\partial k_\mu} \right| m \right\rangle &=\frac{\partial}{\partial k_\mu}\langle n | H | m\rangle -\bigg( \frac{\partial }{\partial k_\mu} \langle n | \bigg) H | m \rangle -\langle n | H \bigg(\frac{\partial}{\partial k_\mu}|m\rangle\bigg) \nonumber \\
&=\frac{\partial}{\partial k_\mu} E_{nm} +\bigg\langle n \bigg| \tau_z \frac{\partial}{\partial k_\mu} \bigg|m \bigg\rangle \left( (\tau_z E)_{mm}-(\tau_z E)_{nn} \right),
\end{align}
where we have used $\langle n|H|m\rangle=E_{nm}$ and 
$\frac{\partial}{\partial k_\mu} \langle n| \tau_z | m \rangle =0$. 
Then, Eq.~\eqref{eq.approx_spin_nernst} can be rewritten as
\begin{equation}
\frac{k_B}{\hbar} \frac{i}{2V} \sum_{\bm{k},m\neq n} \bigg\langle n \bigg| S^z \tau_z \frac{\partial H}{\partial k_\mu}+\frac{\partial H}{\partial k_\mu}\tau_z S^z \bigg|m\bigg\rangle \bigg\langle m\bigg| \tau_z \frac{\partial}{\partial k_\nu} \bigg| n\bigg\rangle (\tau_z)_{nn}(\tau_z)_{mm} \frac{c_1(\frac{E_{mm}}{k_B T})-c_1(\frac{E_{nn}}{k_B T})}{(\tau_z E)_{nn}-(\tau_z E)_{mm}}.
\end{equation}
Since $r_\mu=i\frac{\partial}{\partial k_\mu}$, this expression can be compared with the expression containing $C_S$. 
Using the completeness relation in Eq.~\eqref{eq.completeness}, we have
\begin{equation}
H=\sum_{p} \tau_z |p\rangle E_p \langle p | \tau_z, \quad C_S=\sum_{p,q}\tau_z |p\rangle (\tau_z)_{pp} \langle p| C_S |q\rangle (\tau_z)_{qq} \langle q | \tau_z
\end{equation}
If we consider the energy scale in which the anticrossing between magnon and phonon bands occur, we expect $\langle p | C_S | q\rangle$ to smaller by a factor of 10 for the band crossing with the lowest energy, which occurs around $1$ meV, and we expect it to be much smaller for the band crossing which occurs at higher energies.

\section{Spin Density}

We assume periodicity along the direction of temperature gradient, but assume finite size along the edge perpendicular to it.
Let us first examine the intrinsic contribution to the spin density.
From the Kubo formula, the intrinsic contribution to the spin density is
\begin{equation}
\langle \delta \mathcal{S}^z(r) \rangle=\langle \mathcal{S}^z(r) \rangle_{\textrm{neq}}-\langle \mathcal{S}^z(r) \rangle_{\textrm{eq}}=-\lim_{\omega\rightarrow 0} \frac{\partial}{\partial \omega} \int_0^\beta d\tau e^{i\omega \tau} \langle T_\tau \mathcal{S}^z(r,\tau) J_\nu^Q(0)\rangle_{\textrm{eq}} \nabla_\nu \chi.
\end{equation}
Here, $\mathcal{S}^z(r)$ is the spin density of the $r$th strip, where $r$ is the index for the unit cells along the finite size, which is $x$ for the armchair edge and $y$ for the zigzag edge, as shown in Fig.~\ref{fig.spin_density}.
Taking the Fourier transformation along the periodic direction $\nu$, the BdG fields are $\Psi_{\bm{k},m}$ where $\bm{k}$ is the momentum along the periodic direction and $m=1, ..., 24N$, which contains the index for the magnon and phonon modes contained in a unit cell ($24N$ for both zigzag and armchair edge) multiplied by the number of strips ($N$) .
Let us define
\begin{equation}
\mathcal{S}^z(r)=\sum_{\bm{k}}\Psi^\dagger_{\bm{k}} \frac{S^z_r}{2} \Psi_{\bm{k}}, \quad
\bm{J}^Q=\frac{1}{4}\int d\bm{r} \Psi^\dagger_{\bm{k}} (h_0 \tau_z \bm{v}_{\bm{k}}+\bm{v}_{\bm{k}} \tau_z h_0 )\Psi_{\bm{k}},
\end{equation}
where $\mathcal{S}^z(r)$ is the spin density operator for the $r$th strip. 
Proceeding as in Sec.~\ref{sec.evaluation_kubo}, we have $\langle \delta\mathcal{S}^z(r) \rangle=-Z^{\textrm{in}}_\nu(r) \nabla_\nu \chi$, where
\begin{align}
Z^{\textrm{in}}_\nu(r)&=-\frac{i}{2}\sum_{\bm{k},m\neq n} [T^\dagger_{\bm{k}} S^z_r T_{\bm{k}}]_{mn}[\tau_z (T_{\bm{k}}^\dagger v_{\bm{k},\nu} T_{\bm{k}} \tau_z E_{\bm{k}} + E_{\bm{k}} \tau_zT_{\bm{k}}^\dagger v_{\bm{k},\nu}T_{\bm{k}})\tau_z]_{nm}\frac{g[(\tau_z E_{\bm{k}})_{mm}]-g[(\tau_z E_{\bm{k}})_{nn}]}{[(\tau_z E_{\bm{k}})_{mm}-(\tau_z E_{\bm{k}})_{nn}]^2} \nonumber \\
&=-\frac{i}{2} \sum_{\bm{k},m\neq n} \langle m,\bm{k}|S^z_r | n,\bm{k}\rangle \langle n ,\bm{k}| v_{\bm{k},\nu}|m ,\bm{k}\rangle [(\tau_z)_{nn} (E_{\bm{k}})_{mm}+(\tau_z)_{mm}(E_{\bm{k}})_{nn}] \frac{g[(\tau_z E_{\bm{k}})_{mm}]-g[(\tau_z E_{\bm{k}})_{nn}]}{[(\tau_z E_{\bm{k}})_{mm}-(\tau_z E_{\bm{k}})_{nn}]^2} ,
\end{align}
and the superscript `$\textrm{in}$' indicates the intrinsic contribution. 
Then, $\langle \delta\mathcal{S}^z(r) \rangle=-\zeta^{\textrm{in}}_\nu(r) \nabla_\nu T$, where $\zeta^{\textrm{in}}_\nu(r)=Z^{\textrm{in}}_\nu(r)/T$.

\begin{figure*}[t]
\centering
\includegraphics[width=17cm]{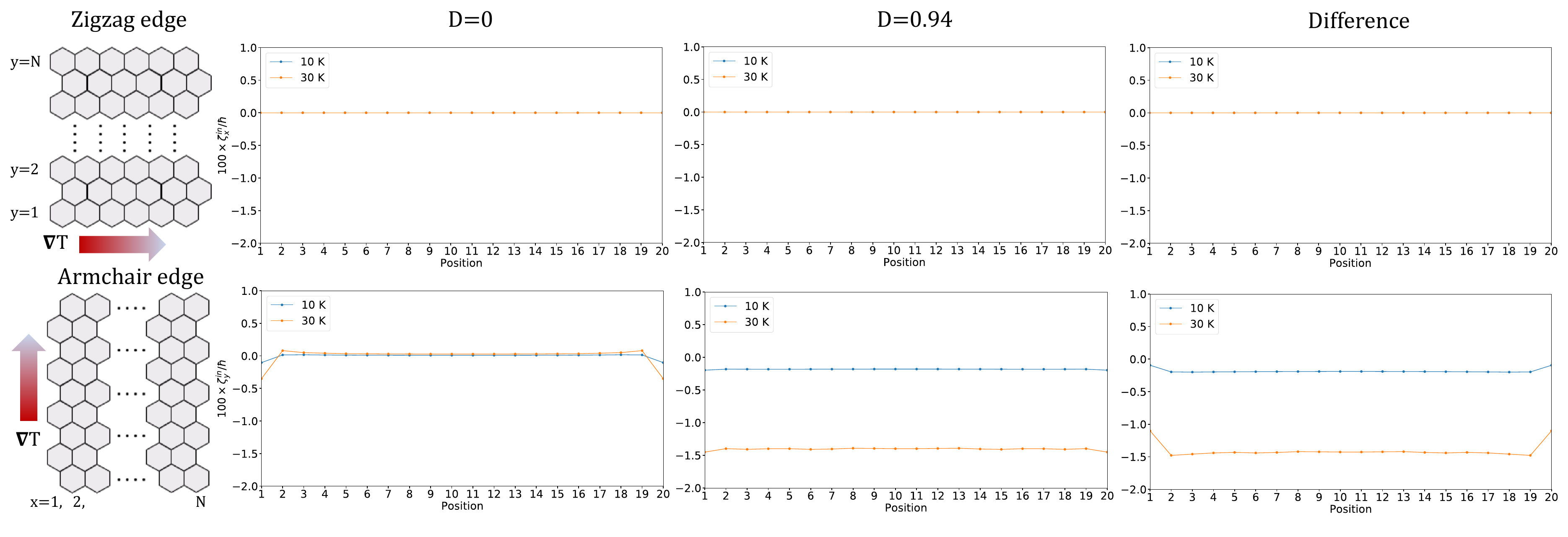}
\caption{ On the left, we show the zigzag and armchair edge configuration for which the intrinsic contribution to the spin density $\zeta^{\textrm{in}}_\nu(r)$ is calculated, which is shown to the right. For reference, we show $\zeta^{\textrm{in}}_\nu(r)$ for the case where there is no magnon-phonon interaction, i.e. $D=0$ meV, so that spin is conserved. We see that for the zigzag edge, no spin density is induced while for the armchair edge, spin density is induced symmetrically at the edges, and which are cancelled by the bulk.  In the presence of magnon-phonon interaction ($D=0.94$ meV), the spin density still vanishes for the zigzag edge, while a relatively large bulk spin density is induced for the armchair edge. The difference $\zeta^{\textrm{in}}_\nu(r)|_{D=0.94}-\zeta^{\textrm{in}}_\nu(r)|_{D=0}$ represents the spin density induced by magnon-phonon interaction.
}
\label{fig.spin_density}
\end{figure*}

In Fig.~\ref{fig.spin_density}, we plot $\zeta^{\textrm{in}}_{\nu}(r)$ for the zigzag and armchair edges, with and without magnon-phonon interaction.
Here, we note that we have put $l=1$, where $l$ is the distance between $A$ and $B$ sites.
Thus, the distance between the unit cells along the $x$ ($y$) direction for the zigzag (armchair) edge is $\sqrt{3}$ ($3$).
As can be seen in Fig.~\ref{fig.spin_density}, in the absence of magnon-phonon interaction, the spin density for the zigzag edge vanishes while the spin density for the armchair edge is almost symmetric.
Such distribution of spin density can be explained by examining the symmetry transformation property of $\zeta^{\textrm{in}}_\nu(r)$.
Even in the presence of the magnon-phonon interaction, the spin density for the zigzag edge remains zero, while the spin density for the armchair edge remains (nearly) symmetric.
To explain such a distribution, we examine some of the properties of the spin density induced by thermal gradient.

First, let us assume that there is no magnon-phonon interaction.
In such a case, the total spin density of the system vanishes.
This occurs for a bipartite antiferromagnetic system whenever the spin is conserved, so that the Hamiltonian can be written in a block form for each spin sector.
For our system (applicable for both the finite size and the bulk system), the magnon Hamiltonian can be written in the following block form,
\begin{equation}
\mathcal{H}=\frac{1}{2}\sum_{\bm{k}}\phi_{\bm{k}}^\dagger H_{\bm{k}} \phi_{\bm{k}}, \quad\phi_{\bm{k}}=\begin{pmatrix}
\phi_{A\bm{k}} \\
\phi_{B-\bm{k}}^{\dagger} \\
\hline
\phi_{B\bm{k}} \\
\phi^\dagger_{A-\bm{k}}
\end{pmatrix}, \quad H_{\bm{k}}=\begin{pmatrix}
H_{I\bm{k}} &\vline& 0 \\
\hline
0 &\vline &H_{II\bm{k}}
\end{pmatrix},
\end{equation}
where $\phi_{A\bm{k}}$ ($\phi_{B\bm{k}}$) contains only the magnon annihilation operators on $A$ ($B$) sites.
In this basis, the total spin density operator is
\begin{equation}
\mathcal{S}^z_{\textrm{tot}}=\frac{1}{2}\sum_{\bm{k}} \Psi^\dagger_{\bm{k}} S^z_{\textrm{tot}}\Psi_{\bm{k}}, \quad S^z_{\textrm{tot}}=\frac{\hbar}{V} \begin{pmatrix}
-1_{2N} & 0 & \vline & 0 & 0 \\
0 & 1_{2N} & \vline & 0 & 0 \\
\hline
0 & 0 & \vline & 1_{2N} & 0 \\
0 & 0 & \vline & 0 & -1_{2N}
\end{pmatrix},
\end{equation}
where $1_{2N}$ is the $2N$ by $2N$ identity matrix.
Because each of the blocks can be diagonalized using paraunitary transformation \cite{colpa1978diagonalization} (see, for example, Refs.~\cite{cheng2016spin,zyuzin2016magnon}), and because $S^z_{\textrm{tot}}$ in each of the blocks is proportional to $\tau_z^{4N}$ where the superscript $4N$ indicates the matrix size, $\langle m,\bm{k}| S_{\textrm{tot}}^z |n,\bm{k}\rangle \propto (\tau_z^{4N})_{mn}$, where we have used the identity $\langle m,\bm{k}| \tau_z^{4N} |n,\bm{k}\rangle = (\tau_z^{4N})_{mn}$.
Since $(\zeta^{\textrm{in}}_{\nu})^{\textrm{tot}}=\sum_{r}\zeta^{\textrm{in}}_{\nu}(r)$ only contains summation over the band indices with $m\neq n$, $(\zeta^{\textrm{in}}_{\nu})^{\textrm{tot}}=0$.
We note that the same derivation goes through whenever the spin is conserved, since the matrix elements of the spin operator is diagonal.
Indeed, it can be checked for the armchair edge that the total spin density vanishes when $D=0$.
However, as can be seen in  the case for the armchair edge, the local spin density need not be zero.
Therefore, we conclude that although the bulk spin density vanishes in the thermodynamic limit, for a finite size system, temperature gradient can induce small spin density in the bulk and comparably large spin density at the edge in such a way that their sum is zero.

Next, let us discuss the constraint to $\zeta^{\textrm{in}}_\nu$ from the $\mathcal{M}_x\mathcal{C}_{2x}^S$ symmetry introduced in the main text in the presence of edges.
Here, $\mathcal{M}_{x}$ is the mirror symmetry about the plane normal to the $x$ axis and passing through one of the lines connecting $A$ and $B$ sites, and $\mathcal{C}_{2x}^S$ is a twofold rotation symmetry about the $x$ axis which acts only on the spin degrees of freedom and does not act on their position. 
This symmetry is present when we neglect the magnon-phonon interaction (i.e. $D=0$).
Letting $M_x^S$ be the matrix representation of this symmetry, we find that 
in the case of zigzag edge, $(M_x^S)^\dagger H_{k_x}(M_x^S)=H_{-k_x}$, $(M_x^S)^\dagger S_y^z M_x^S=S_y^z$, and $M_x^S|n,k_x\rangle =|n,-k_x\rangle$, from which it follows that $\zeta^{\textrm{in}}_x(y)=-\zeta^{\textrm{in}}_x(y)=0$.
Specifically, if we define $f((E_{k_x})_{mm},(E_{k_x})_{nn})=[(\tau_z)_{nn} (E_{k_x})_{mm}+(\tau_z)_{mm}(E_{k_x})_{nn}] \frac{g[(\tau_z E_{k_x})_{mm}]-g[(\tau_z E_{k_x})_{nn}]}{[(\tau_z E_{k_x})_{mm}-(\tau_z E_{k_x})_{nn}]^2}$, we have
\begin{align}
Z^{\textrm{in}}_x(y)&=-\frac{i}{2} \sum_{k_x,m\neq n} \langle m,k_x|S^z_y | n,k_x\rangle \langle n ,k_x| \partial_x H_{k_x}|m ,k_x\rangle f((E_{k_x})_{mm},(E_{k_x})_{nn})  \nonumber \\
&=-\frac{i}{2} \sum_{k_x,m\neq n} \langle m,-k_x|S^z_y | n,-k_x\rangle \langle n ,-k_x| \partial_x H_{-k_x}|m ,-k_x\rangle f((E_{-k_x})_{mm},(E_{-k_x})_{nn}) \nonumber \\
&=-Z^{\textrm{in}}_x(y).
\end{align}
For the armchair edge, if we neglect the slight asymmetry from the edge configuration, the lattice is symmetric under $\mathcal{M}_x$ passing through the center of the lattice. Letting $M_x x$ denote position of the position $x$ under the action of this symmetry, we have $(M_x^S)^\dagger H_{k_y}(M_x^S)=H_{k_y}$, $(M_x^S)^\dagger S_{x}^z M_x^S=S_{M_x x}^z$, and $M_x^S|n,k_y\rangle =|n,k_y\rangle$, from which it follows that $\zeta^{\textrm{in}}_y(x)=\zeta^{\textrm{in}}_y(M_x x)$.
Thus, the symmetry constrains the distribution of spin density to be symmetric along the $x$ axis, as can be checked in Fig.~\ref{fig.spin_density}.

Next, let us show that the $\mathcal{C}_{3z}$ symmetry, which is present even when $D\neq 0$, constrains $\zeta^{\textrm{in}}_\nu=0$ for periodic system (i.e. system without edges).
To derive this, let $C_{3z}$ be the representation of the symmetry, i.e. $C_{3z}^\dagger H_{\bm{k}}C_{3z}=H_{C_{3z}\bm{k}}$ and $|n,\bm{k}\rangle=C_{3z}|n C_{3z}\bm{k}\rangle$.
We find that $\bm{\zeta^{\textrm{in}}}=C_{3z}\bm{\zeta^{\textrm{in}}}=0$. 
That is, threefold rotation symmetry disallows bulk spin density induced by thermal gradient.
Indeed, we find that there is no bulk spin density in our system regardless of the presence of magnon-phonon interaction.
On the other hand, this does not require vanishing spin density in the presence of zigzag and armchair edges since the threefold symmetry is broken by the presence of edges.
Before moving on, let us note that the inversion symmetry also constrains $\bm{\zeta}^{\textrm{in}}=0$.

\begin{figure*}[t]
\centering
\includegraphics[width=17cm]{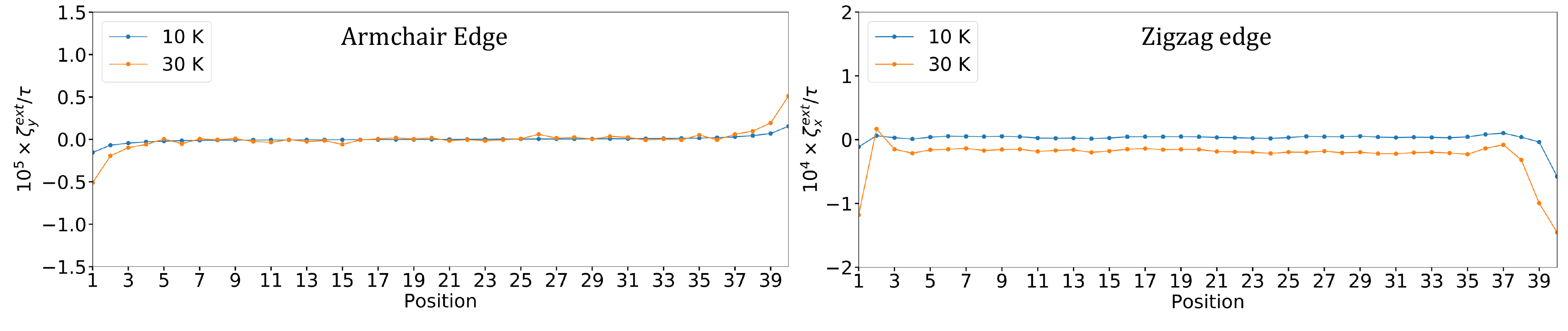}
\caption{ Left: extrinsic contribution to the spin density for the armchair edge when $D=0.94$ meV.
Right: extrinsic contribution to the spin density for the zigzag edge when $D=0.94$ meV.
Note that the spin density induced by the extrinsic mechanism uniformly vanishes when there is no magnon-phonon interaction.
}
\label{fig.spin_density_ext}
\end{figure*}

Next, let us discuss the zigzag and armchair edges when $D\neq 0$.
We find that the spin density for the zigzag edge continues to be identically zero even though the $\mathcal{M}_x \mathcal{C}_{2x}^S$ is broken, and the spin density for the armchair continues to be symmetric along the $x$ axis.
This can be explained using a rather complicated antiunitary symmetry, which we denote by $\mathcal{Y}$.
Let us first discuss its constraints on the bulk properties.
Let $\mathcal{K}$ be the complex conjugation operator and let $\mathcal{m}_{u_y}$ be the operator that sends $u_{y}(r)\rightarrow -u_{y}(r)$ while keeping everything else fixed, where $u_{y}(r)$ is the displacement operator in the $y$ direction located at position $r$.
Let $\mathcal{m}_x$ be the operator that sends $\Psi(r)\rightarrow \Psi({M_x r})$, where $\Psi(r)$ is the field operator that contains both the magnon and phonon fields located position $r$.
Then, we define $\mathcal{Y}\equiv \mathcal{K} \mathcal{m}_{u_y} \mathcal{m}_x$.
Roughly, $\mathcal{Y}$ can be understood as a combination of pseudo time reversal symmetry $\mathcal{K}$ which does not flip the spin direction, with the pseudo mirror symmetry $\mathcal{m}_{u_y}\mathcal{m}_x$ about the plane normal to the $x$ axis which also does not flip the spin direction. 
Let $Y\mathcal{K}$ be the matrix representation of $\mathcal{Y}$ in the $\bm{k}$ space. 
We note that $Y$ is the matrix that sends $u_{A/B\bm{k}}^y \rightarrow -u_{A/B\bm{k}}^y $ and act as an identity on other field operators.
Under this symmetry, we find that $Y^\dagger H^*_{(k_x,k_y)}Y=H_{(k_x,-k_y)}$ and $|n, k_x,-k_y \rangle=Y^\dagger \mathcal{K}|n,k_x,k_y \rangle$.
From this, it follows that $\Omega_{n,(k_x,k_y)}\rightarrow \Omega_{n,(k_x,-k_y)}$ and $\langle \mathcal{S}^z\rangle_{n,(k_x,k_y)}\rightarrow \langle \mathcal{S}^z\rangle_{n,(k_x,-k_y)}$, the point being that the thermal Hall and spin Nernst currents are not forbidden.
On the other hand, proceeding similarly as before, we find that $\zeta^{\textrm{in}}_{x}(y)\rightarrow -\zeta^{\textrm{in}}_{x}(y)=0$ for the zigzag edge and $\zeta^{\textrm{in}}_{y}(x)\rightarrow \zeta^{\textrm{in}}_{y}(M_x x)$ for the armchair edge (if we ignore the slight asymmetry caused by the edge configureation).
\textit{We therefore conclude that because of the $\mathcal{Y}$ symmetry, the intrinsic contribution to the spin density does not cause asymmetric spin distribution.} 

Next, let us evaluate the extrinsic contribution to the spin density. 
The Boltzmann transport theory within the constant relaxation time approximation gives \cite{ashcroft1976solid} $g_{\textrm{neq}}(E)= g_{\textrm{eq}}(E)-\tau v_{\nu}\nabla T_\nu \frac{E}{k_BT^2}\frac{e^{E/k_BT}}{(e^{E/k_BT}-1)^2}$, so that
\begin{equation}
\zeta^{\textrm{ext}}_{\nu}(r)=\frac{\tau}{2k_BT^2}\frac{1}{V} \sum_{\bm{k}}\sum_{n=-N}^{N} \langle n,\bm{k}| S_r^z | n,\bm{k} \rangle \langle n,\bm{k}| v_{\bm{k},\nu}|n,\bm{k}\rangle (\tau_z E_{\bm{k}})_{nn} \frac{e^{(\tau_z E_{\bm{k}})_{nn}/k_BT}}{(e^{(\tau_z E_{\bm{k}})_{nn}/k_BT}-1)^2}.
\end{equation}
We note that the same equation can be derived using the Kubo formula by making a constant lifetime approximation and restricting to intraband contribution, as in Ref.~\cite{mook2019spin}.
As discussed in the case for intrinsic contribution to the spin density, when the system is symmetric with respect to $\mathcal{M}_x \mathcal{C}_{2x}$, this symmetry constrains the spin density to appear symmetrically for the armchair edge, and 
constrains the spin density to vanish for the zigzag edge.
On the other hand, the $\mathcal{Y}$ symmetry constrains the spin density to appear antisymmetrically for the armchair edge, but it does not constrain the spin density for the zigzag edge.
Thus, in the absence of magnon-phonon coupling, in which case both symmetries are present, the spin density vanishes for both the armchair and zigzag edge.
However, in the presence of magnon-phonon coupling, only the $\mathcal{Y}$ symmetry is present, so that the spin density appears antisymmetrically for the armchair edge, while no constraint is imposed on the zigzag edge. 
From this, we see that total spin density is forced to vanish in the armchair edge while nonzero total spin density can be induced in the presence of zigzag edge.
Let us note that this behavior of the extrinsic contribution to the spin density is opposite to that of the intrinsic contribution, and that the origin of this behavior lies in the different behavior of intrinsic and extrinsic contributions under the action of antiunitary symmetry, such as $\mathcal{Y}$.
We confirm the above statements numerically as shown in Fig.~\ref{fig.spin_density_ext}, where we show the extrinsic contribution to the spin density in the presence of magnon-phonon interaction.
\textit{We conclude that the intrinsic spin Nernst current induces asymmetric spin density through the extrinsic contribution to the spin density.}

Finally, let us comment on the magnetization induced by temperature gradient. 
According to Refs.~\cite{shitade2019theory,shitade2019gravitomagnetoelectric}, the magnetization susceptibility $\mu_\nu$ which we define from the relation $\langle  \delta \mathcal{M}^z \rangle=-\mu_\nu \nabla_\nu T$, is given by
\begin{equation}
\mu_\nu=\frac{1}{T}\sum_{n,\bm{k}} \textrm{Re}\left[\sum_{m\neq n}\frac{ i\langle n,\bm{k}|\partial_{\nu} H_{\bm{k}}|m,\bm{k}\rangle\langle m,\bm{k}| M^z_{\textrm{tot}} | n,\bm{k} \rangle \tau_{mm} \tau_{nn}}{((\tau_z E_{\bm{k}})_{nn}-(\tau_z E_{\bm{k}})_{mm})^2}\right]\left[(\tau_z E_{\bm{k}})_{nn} g((\tau_z E_{\bm{k}})_{nn}) +\int_{(\tau_z E_{\bm{k}})_{nn}}^\infty dx g(x)\right].
\end{equation}
Here, $ \langle \delta \mathcal{M}^z \rangle=\langle \tilde{\mathcal{M}^z}\rangle  _{\textrm{neq}}-\langle \mathcal{M}^z \rangle_{\textrm{eq}}$ is the change in the magnetization density arising from the temperature gradient, and the quantities $\mathcal{M}^z$ and $\tilde{\mathcal{M}}^z$ are the equilibrium and non-equilibrium magnetization density, respectively, and are defined as
\begin{equation}
\mathcal{M}^z=\frac{1}{2} \sum_{\bm{k}} \Psi_{\bm{k}}^\dagger M^z \Psi_{\bm{k}}, \quad \tilde{\mathcal{M}}^z=\frac{1}{2} \sum_{\bm{k}} \tilde{\Psi}_{\bm{k}}^\dagger M^z \tilde{\Psi}_{\bm{r}}, \quad M^z_{\textrm{tot}}=\frac{1}{V} \begin{pmatrix}
-\mu_A \times1_{2N} & 0 & \vline & 0 & 0 \\
0 & \mu_B \times1_{2N} & \vline & 0 & 0 \\
\hline
0 & 0 & \vline & \mu_B\times 1_{2N} & 0 \\
0 & 0 & \vline & 0 & -\mu_A \times1_{2N}
\end{pmatrix},
\end{equation}
where $\mu_A$ and $\mu_B$ are the magnetic moment of the spins at $A$ and $B$ sites (for the definition of $\tilde{\Psi}$, see Eq.~\eqref{eq.tilde_psi}).
We note that this quantity is not derived from the $\zeta^{\textrm{in}}_\nu$ by simply replacing $S^z_{\textrm{tot}}$ with $M^z_{\textrm{tot}}$ because the definition of magnetization is modified from $\mathcal{M}^z$ to $\tilde{\mathcal{M}}^z$ in the presence of temperature gradient \cite{shitade2019theory,shitade2019gravitomagnetoelectric}. 
We note that the properties of the spin density operator we have discussed can be straightforwardly applied to this quantity as well. 
In particular, the symmetry constraints are the same when $\mu_A=\mu_B$.

\end{widetext}

\end{document}